\documentclass[twocolumn]{jpsj2}
\usepackage{graphicx}
\usepackage{amsmath,amssymb}

\title{ 
Magnetic Properties in 
Non-centrosymmetric Superconductors 
\\ with and without Antiferromagnetic Order
} 

\author{Youichi {\sc Yanase}$^{1,2}$\footnote{E-mail:
yanase@itp.phys.ethz.ch} and Manfred {\sc Sigrist}$^{2,3}$}

\inst{1 Department of Physics, University of Tokyo, Tokyo 113-0033, Japan \\
\noindent 2 Theoretische Physik, ETH-Honggerberg, 8093 Zurich, Switzerland \\
\noindent 3 Department of Physics, Kyoto University, Kyoto 606-8502, Japan}

\recdate{Today 2007}

\abst
{ 
 The paramagnetic properties in non-centrosymmetric superconductors with 
and without antiferromagnetic (AFM) order are investigated with focus on 
the heavy Fermion superconductors, CePt$_3$Si, CeRhSi$_3$ and CeIrSi$_3$. 
 First, we investigate the spin susceptibility in the linear response regime 
and elucidate the role of AFM order. 
 The spin susceptibility at $T=0$ is independent of the pairing symmetry 
and increases in the AFM state. 
 Second, the non-linear response to the magnetic field are investigated 
on the basis of an effective model for CePt$_3$Si which may be also 
applicable to CeRhSi$_3$ and CeIrSi$_3$. 
 The role of antisymmetric spin-orbit coupling (ASOC), 
helical superconductivity, anisotropic Fermi surfaces 
and AFM order are examined in the dominantly $s$-, $p$- and $d$-wave states. 
We emphasize the qualitatively important role of the mixing of superconducting 
(SC) order parameters in the $p$-wave state  which enhances the 
spin susceptibility and suppresses paramagnetic depairing effect in a significant way. 
 Therefore, the dominantly $p$-wave superconductivity admixed with 
the $s$-wave order parameter is consistent with the paramagnetic 
properties of CePt$_3$Si at ambient pressure. 
 We propose some  experiments which can elucidate the novel 
pairing states in CePt$_3$Si as well as CeRhSi$_3$ and CeIrSi$_3$. 
}

\kword
{
Superconductivity without inversion center; antiferromagnetic superconductor;
Pauli paramagnetic effect; spin susceptibility
}

\begin{document}
\sloppy
\maketitle

\newcommand{\eli}{$\acute{{\rm E}}$liashberg }
\renewcommand{\k}{\vec{k}}
\newcommand{\kp}{\vec{k}_{+}}
\newcommand{\km}{\vec{k}_{-}}
\newcommand{\kk}{\vec{k}'}
\newcommand{\kkk}{\vec{k}''}
\newcommand{\kpk}{\vec{k}'_{+}}
\newcommand{\kmk}{\vec{k}'_{-}}
\newcommand{\q}{\vec{q}}
\newcommand{\qfflo}{\vec{q}_{\rm H}}
\newcommand{\Q}{\vec{Q}}
\newcommand{\h}{\vec{h}}
\newcommand{\e}{\varepsilon}
\newcommand{\ee}{e}
\newcommand{\s}{{\mit{\it \Sigma}}}
\newcommand{\J}{\mbox{\boldmath$J$}}
\newcommand{\vv}{\mbox{\boldmath$v$}}
\newcommand{\Jh}{J_{{\rm H}}}
\newcommand{\LL}{\mbox{\boldmath$L$}}
\renewcommand{\SS}{\mbox{\boldmath$S$}}
\newcommand{\Tc}{$T_{\rm c}$ }
\newcommand{\Tcf}{$T_{\rm c}$}
\newcommand{\etal}{{\it et al.}: }
\newcommand{\Co}{${\rm Na_{x}Co_{}O_{2}} \cdot y{\rm H}_{2}{\rm O}$ }
\newcommand{\Cof}{${\rm Na_{x}Co_{}O_{2}} \cdot y{\rm H}_{2}{\rm O}$}
\newcommand{\hc}{$h_{\rm c2}$ }
\newcommand{\hcf}{$h_{\rm c2}$}
\newcommand{\Hc}{$H_{\rm c2}$ }
\newcommand{\Hcf}{$H_{\rm c2}$}
\newcommand{\Pt}{CePt$_3$Si }
\newcommand{\Rh}{CeRhSi$_3$ }
\newcommand{\Ir}{CeIrSi$_3$ }
\newcommand{\Ptf}{CePt$_3$Si}
\newcommand{\Rhf}{CeRhSi$_3$}
\newcommand{\Irf}{CeIrSi$_3$}
\newcommand{\PRL}{Phys. Rev. Lett. }
\newcommand{\PRB}{Phys. Rev. B }
\newcommand{\JPSJ}{J. Phys. Soc. Jpn. }
\newcommand{\Onuki}{${\rm{\bar O}}$nuki}
\newcommand{\Onukis}{${\rm{\bar O}}$nuki }

\section{Introduction}

Since the discovery of superconductivity in the non-centrosymmetric 
heavy Fermion compound
CePt$_3$Si,~\cite{rf:bauerDC,rf:bauerreview} 
superconductivity in materials without inversion center is 
attracting growing interest. 
Many new non-centrosymmetric superconductors (NCSC) with unusual properties 
have been identified among heavy fermion systems 
such as UIr,~\cite{rf:akazawa} CeRhSi$_3$,~\cite{rf:kimura,rf:kimurareview} 
CeIrSi$_3$,~\cite{rf:sugitani,rf:onukireview} 
CeCoGe$_3$~\cite{rf:CeCoGe} and others like
Li$_2$Pd$_{x}$Pt$_{3-x}$B,~\cite{rf:togano} 
Y$_2$C$_3$,~\cite{rf:akimitsu}
Rh$_2$Ga$_9$, Ir$_2$Ga$_9$,~\cite{rf:shibayama,rf:akimitsuprivate} 
Mg$_{10}$Ir$_{19}$B$_{16}$,~\cite{rf:mu} 
Re$_3$W~\cite{rf:zuev} 
and some organic materials.~\cite{rf:ohmichi}
The aspects of missing inversion symmetry are also 
of great interest for other materials. 
For example, the spin Hall effect in the semiconductor~\cite{rf:spinHall} 
and the helical magnetism in MnSi~\cite{rf:MnSi} 
are very active research fields.

NCSC adds several unusual aspects to the properties of superconductivity.
One immediate consequence of non-centrosymmetricity is
the necessity for an extended classification scheme of Cooper pairing 
states, as parity is not available as a distinguishing symmetry. Using the
traditional scheme the SC states here 
may be represented as a mixture of pairing states of even and odd parity, 
or, equivalently, their spin configuration is a superposition of
a singlet and a triplet component. This is a consequence of 
the presence of antisymmetric spin-orbit coupling 
(ASOC) in non-centrosymmetric materials.~\cite{rf:edelsteinMixChi}  
Recent theoretical studies led to the discussion of various
intriguing properties which could appear in NCSC, such as the 
magneto-electric effect,~\cite{rf:edelsteinMixChi,
rf:edelsteinMEE,rf:fujimotoChi,rf:fujimoto,rf:yip} 
the unusual anisotropic spin susceptibility,~\cite{rf:edelsteinMixChi,
rf:yip,rf:fujimotoChi,rf:samokhinChi,rf:gorkov,rf:frigeri,rf:frigeriChi,
rf:yanaselett,rf:bulaevskii} the occurrence of an
anomalous coherence effect in NMR $1/T_{1}T$,~\cite{rf:fujimoto,rf:hayashiT1} 
the unusual origin of nodes in the 
SC gap,~\cite{rf:fujimotoGap,rf:hayashiSD,rf:eremin,
rf:yanaselett,rf:hayashiT1} 
the realization of the helical SC phase,~\cite{rf:samokhinHelical,
rf:kaurHV,rf:agterberg,rf:oka,rf:mineevHelical,rf:feigelman} 
the possible appearance of Fulde-Ferrel-Larkin-Ovchinnikov (FFLO) state 
at zero magnetic field,~\cite{rf:tanaka} 
de Haas-van Alphen effect,~\cite{rf:mineevdHvA}
various novel impurity 
effects,~\cite{rf:mineevImp,rf:frigeriImp,rf:samokhinImp} and 
vortex core states~\cite{rf:nagai} 
and unconventional features in quasiparticle tunneling and 
Josephson effect.~\cite{rf:yokoyama,rf:sudbo,rf:linder,rf:hayashiJP,
rf:sergienko,rf:iniotakis,rf:varma,rf:borkje}

 The non-centrosymmetric heavy fermion superconductors, e.g.  
CePt$_3$Si, UIr, CeRhSi$_3$, CeIrSi$_3$ and CeCoGe$_3$ are of 
particular interests because the Cooper pairing is most likely 
unconventional (non-$s$-wave) due to the strong electron correlation. 
 Although many studies have been devoted to this topics, 
there is no consensus 
on the symmetry of pairing in these compounds so far. 
 The symmetry of Cooper pairs may be determined by the paramagnetic 
properties such as the spin susceptibility below \Tcf.

 In centrosymmetric superconductor, the spin susceptibility is 
a distinguishing feature for
 the spin configuration of the pairing state, as it decreases below \Tc 
for the spin singlet superconductor and remains constant in the case of
spin triplet pairing, if the magnetic field is perpendicular to the 
$d$-vector 
(parallel to the equal-spin direction).~\cite{rf:leggett,rf:sigrist} 
 The measurements of the Knight shift which is proportional to the spin 
susceptibility have played an important role for the identification 
of SC state in various compounds.~\cite{rf:tou-ishida} For superconductors 
with very high \Hc probing effects of paramagnetic limiting can give also 
insight into the pairing symmetry and has been applied in connection 
with NCSC. 
However, the response to the magnetic field is not so straightforward in 
non-centrosymmetric systems. As mentioned above, spin singlet and triplet 
components 
are mixed in the pairing state. Furthermore the band splitting induced 
by the ASOC affects the magnetic 
properties. Therefore, 
it is necessary to clarify the magnetic properties very carefully
before drawing strong conclusions.
In this context also the influence of AFM order on
the (magnetic) properties of the SC phase is an important
point to investigate, 
since all presently known non-centrosymmetric heavy Fermion 
superconductors, i.e. CePt$_3$Si, UIr, CeRhSi$_3$, CeIrSi$_3$ and 
CeCoGe$_3$, coexist with the magnetism.
 In CePt$_3$Si at ambient pressure, superconductivity 
($T_{\rm c}=0.75$K) coexists with AFM order 
($T_{\rm N}=2.2$K).~\cite{rf:bauerDC,rf:amatoCePtSi,
rf:metoki} 
 The AFM order can be suppressed by pressure and vanishing at 
the critical value of $P \sim 0.6$GPa. The SC phase 
is more robust and a purely SC 
phase exists beyond the critical pressure, 
($P > 0.6$GPa).~\cite{rf:yasuda,rf:tateiwa,rf:takeuchiP} 
 CeRhSi$_3$,~\cite{rf:kimura,rf:kimurareview} 
CeIrSi$_3$~\cite{rf:sugitani,rf:onukireview} 
and CeCoGe$_3$~\cite{rf:CeCoGe} are AFM 
at ambient pressure and superconductivity appears only under 
substantial pressure. 
 Although most of the theoretical studies except for Refs.~27, 30 and 61 
neglected the AFM order so far, it turns out that the magnetism affects the electronic state 
profoundly. 
 It has been shown that a gap line-node behavior could be induced by the 
AFM order for the pairing state with dominantly $p$-wave 
component,~\cite{rf:fujimotoGap,rf:yanaselett,rf:yanasefull} 
which may explain the experimental results in CePt$_3$Si at ambient 
pressure.~\cite{rf:izawa,rf:bonalde,rf:takeuchiC} 

In this paper we investigate the linear as well as the non-linear response regime 
of the NCSC in a magnetic field with the aim to provide guidelines to identify
the pairing symmetry based on magnetic properties. Before going into 
details we briefly summarize the main conclusions of our study. 
It is known that in the linear response regime the 
paramagnetic properties are {\it universal} i.e. the spin 
susceptibility is independent of the pairing symmetry. 
 In the presence of Rashba-type ASOC, the spin susceptibility along the 
{\it c}-axis is constant through \Tc while it decreases along the {\it ab}-plane 
to half of the normal state value at $T=0$, in absence of AFM order.~\cite{rf:edelsteinMixChi,rf:fujimotoChi,
rf:samokhinChi,rf:gorkov,rf:frigeri,rf:frigeriChi,rf:yanaselett,rf:bulaevskii} 
 The influence of helicity (Cooper pairs possess a finite momentum) 
 in NCSC on the behavior of the susceptibility turns out to be negligible. 
 On the other hand, the folding of Brillouin zone due to the AFM order 
significantly affects the spin susceptibility in the SC phase.
 The spin susceptibility for the magnetic field perpendicular
(parallel) to the staggered moment is increased (decreased) by the AFM 
order.~\cite{rf:yanaselett}

 The non-linear response to the magnetic field is important 
when the magnetic field is comparable to or higher than the 
standard paramagnetic limiting field
$H_{\rm P} \sim 1.2 k_{\rm B}T_{\rm c}/\mu_{\rm B}$. 
 It should be noted that most of the experimental studies, such as 
the Knight shift and critical magnetic field \Hcf, 
have been carried out 
in the non-linear response region.~\cite{rf:bauerDC,rf:bauerreview,
rf:onukireview,rf:sugitani,rf:kimura,rf:kimurareview,rf:yasuda,rf:yogiK,
rf:higemoto,rf:okuda,rf:mukudaprivate,rf:kimuraprivate} 
 The pairing state in NCSC can be 
identified by the measurements in the non-linear response regime
because the paramagnetic properties depend on the pairing symmetry 
in contrast to the situation in the linear response regime.

 We show that the critical magnetic field \Hc along the {\it ab}-plane 
is significantly enhanced in the non-linear response regime by 
the formation of helical SC state. This enhancement coincides with the 
non-linear increase of the helicity of the SC order parameter. 
\Hc furthermore rises for the dominantly $p$-wave state 
owing to the mixing of SC order parameters. 
 These effects, namely (i) the formation of the helical SC state and (ii) 
the mixing of SC order parameters, are quantitatively important for 
anisotropic Fermi surfaces. AFM order significantly enhances 
the effect (ii) and also boosts \Hcf. In this case,   
the spin susceptibility remains nearly constant through \Tcf. 
On the other  hand, these effects are 
negligible in the dominantly spin 
singlet pairing state. 
 Since the influence of AFM order is quantitatively important, 
the paramagnetic properties of the SC phase in the AFM state 
provide a means to distinguish between pairing states with
dominant spin triplet and singlet component.

 Among the non-centrosymmetric heavy fermion superconductors, 
\Pt has been investigated in most detail 
because the superconductivity exists at ambient pressure 
while others require substantial pressure to become superconducting. 
 Therefore, we pay particular attention to the situation in \Ptf, and 
discuss the pairing symmetry by comparing the 
experiments~\cite{rf:yogiK,rf:higemoto,rf:bauerDC,
rf:bauerreview,rf:onukireview,rf:yasuda} with our theoretical results. 
 The paramagnetic properties of \Pt look puzzling at first sight
because the experimental results are incompatible with the 
theoretical results within the 
linear response theory and without taking into account the AFM 
order.~\cite{rf:edelsteinMixChi,rf:fujimotoChi,rf:samokhinChi,
rf:gorkov,rf:frigeri,rf:frigeriChi,rf:bulaevskii} 
In our present study we show that the experimental results are consistent with 
the theoretical results for the dominantly $p$-wave state by taking 
into account the AFM order as well as the non-linear response to the
magnetic field.

 Moreover we propose further test experiments which could strengthen 
our conclusions.
 First, the influence of AFM order can be examined by the pressure 
which suppresses the AFM order. 
 Second, the 2-fold anisotropy in the {\it ab}-plane arises from the 
AFM order and the anisotropy is qualitatively different between 
the dominantly $p$-wave, inter-plane $d$-wave and intra-plane 
$s$- or $d$-wave states.  
Future experimental studies of these kind could help identify the pairing 
symmetry in CePt$_3$Si, CeRhSi$_3$, CeIrSi$_3$ and CeCoGe$_3$.

 The paper is organized as follows. 
 In \S2 we summarize the linear response theory for the paramagnetic 
properties in NCSC. 
 In \S3 we introduce the effective model for CePt$_3$Si which could be also 
applied to CeRhSi$_3$, CeIrSi$_3$ and CeCoGe$_3$.  
 The paramagnetic properties in the magnetic field along 
the {\it ab}-plane are investigated in \S4 and \S5. 
 The non-linear response to the magnetic field for the dominantly $s$-wave 
state is investigated in \S4. 
 In \S5, which is the main part of this paper, we show the magnetic 
properties in the dominantly $p$-wave state. The influences of the 
helical superconductivity, anisotropic Fermi surface and AFM order are 
elucidated. 
 The pairing symmetry of CePt$_3$Si is discussed in \S6 by comparing the 
experimental results with our theoretical results. 
 Some test experiments are proposed for CePt$_3$Si, CeRhSi$_3$ and 
CeIrSi$_3$ in \S7. 
 In \S8, nature of the helical SC state is investigated 
in details. 
 We show the crossover from the helical SC state with long wave length 
to that with short wave length. 
 These results are summarized and some discussions are given in \S9.

\section{Linear Response Theory}

In this section we investigate the linear response regime of the NCSC
in a magnetic field, and study the magnetic properties 
in the paramagnetic (PM) and in the AFM phase. 
The latter we consider both for the case of a centrosymmetric 
and a non-centrosymmetric system. 

\subsection{General spin susceptibility}

In a first step we derive a general expression for the spin 
susceptibility in the SC state on the basis of the extended 
BCS Hamiltonian, given by
\begin{eqnarray}
\label{eq:BCS-model}
&& \hspace*{-5mm}  
H = H_{\rm b} + H_{\rm SO} + H_{\rm AF} + H_{\Delta},
\\ && \hspace*{-5mm} 
H_{\rm b} = \sum_{\k,s} \e(\k) c_{\k,s}^{\dag}c_{\k,s}, 
\\ && \hspace*{-5mm} 
H_{\rm SO} = \alpha  \sum_{\k} \vec{g}(\k) \cdot \vec{S}(\k),
\\ && \hspace*{-5mm}   
H_{\rm AF} = - \sum_{\k} \vec{h}_{\rm Q}  \cdot \vec{S}_{\rm Q}(\k),
\\ && \hspace*{-5mm}   
H_{\Delta} = -\sum_{s,s',\k} 
[\Delta_{1,s,s'}(\k) c_{-\km,s'}^{\dag} c_{\kp,s}^{\dag}
\nonumber \\ && \hspace*{15mm}
+\Delta_{2,s,s'}(\k) c_{-\km+\Q,s'}^{\dag} c_{\kp,s}^{\dag} 
+ h.c.], 
\end{eqnarray}
where $\k_{\pm} = \k \pm \qfflo/2$,  
$\vec{S}(\k)=\sum_{ss'} \vec{\sigma}_{ss'} c_{\k,s}^{\dag}c_{\k,s'}$ and 
$\vec{S}_{\rm Q}(\k)=\sum_{ss'} 
\vec{\sigma}_{ss'} c_{\k+\Q,s}^{\dag}c_{\k,s'}$. 
 Here $\qfflo$ is the total momentum of Cooper pairs. Note that $\qfflo$ is zero 
in the usual BCS state while that is finite in the helical SC state.~\cite{
rf:samokhinHelical,
rf:kaurHV,rf:agterberg,rf:oka,rf:mineevHelical,rf:feigelman}  
In NCSC the helical SC state can be realized under  magnetic field above 
$H_{\rm c1}$.  We consider a tetragonal crystal lattice and assign the {\it x}-, {\it y}- 
and {\it z}-axis to {\it a}-, {\it b}- and {\it c}-axis, respectively. 

The first term in eq.~(1) describes the dispersion relation without 
ASOC and AFM order. In this subsection we 
do not identify the specific dispersion of the electrons and assume 
$\e(\k)$ as general.   

The second term $H_{\rm SO}$ describes the ASOC due to the lack of 
inversion symmetry. This term preserves time reversal symmetry, if the $g$-vector 
is odd in $\k$, i.e. $\vec{g}(-\k)=-\vec{g}(\k)$. We consider a Rashba-type 
spin-orbit coupling~\cite{rf:rashba} as is realized in CePt$_{3}$Si, CeRhSi$_{3}$,  
CeIrSi$_3$ and CeCoGe$_3$.~\cite{rf:frigeri} 
Because the detailed momentum dependence of $\vec{g}(\k)$ is unknown, 
we express it in terms of velocities 
$\vec{v}(\k) = \partial \e(\k)/\partial \k$ : 
$\vec{g}(\k)= (- v_{\rm y}(\k) , v_{\rm x}(\k), 0) /\bar{v}$.  This choice at least
preserves the correct periodicity in $\k$-space. 
The detailed form of the $g$-vector is anyway unimportant in the following.  
We normalize the $g$-vector $\vec{g}(\k)$ by the average velocity $\bar{v}$ 
[$\bar{v}^{2}=\frac{1}{N}\sum_{k}v_{\rm x}(\k)^{2}+v_{\rm y}(\k)^{2}$] 
so that the coupling constant $\alpha$ has the dimension of energy. 
 We assume the relation $|\Delta_{i,s,s'}(\k)| \ll |\alpha| \ll \e_{\rm F}$ 
throughout this paper ($\e_{\rm F}$ is the Fermi energy). 
 This relation is valid for the most of NCSC such as 
CePt$_{3}$Si,  UIr, CeRhSi$_3$, CeIrSi$_3$ and CeCoGe$_3$.

The third term $H_{\rm AF}$ is taken into account to investigate the 
role of AFM order which enters through the staggered field 
$\vec{h}_{\rm Q}$. 
 We focus on A-type AFM order, i.e. ferromagnetic sheets in 
the {\it ab}-plane are staggered along 
the {\it c}-axis, giving rise to $\Q = (0,0,\pi)$. 
This spin structure is realized in 
CePt$_3$Si~\cite{rf:metoki} as well as the centrosymmetric superconductor 
UPd$_2$Al$_3$~\cite{rf:Geibel,rf:Krimmel} where the magnetic moments are
aligned in the {\it ab}-plane. 
 A different AFM state has been reported for \Rhf~\cite{rf:aso} 
and the magnetic structure is not clearly identified for \Ir so far.
 However, the qualitative role of 
AFM order can be captured by $\Q = (0,0,\pi)$  in the simple cases.

The last term $H_{\Delta}$ describes the mean field term of the SC order. 
The order parameter is given by $\Delta_{1,s,s'}(\k)$ and 
$\Delta_{2,s,s'}(\k)$. 
 The second component  $\Delta_{2,s,s'}(\k)$ only appears in the case
 of superconductivity coexisting with AFM order 
($\Delta_{2,s,s'}(\k)=0$ for $\vec{h}_{\rm Q}=0$). 
  The order parameter has both the spin singlet and triplet components 
owing to the ASOC. 

It is more transparent for the following discussion 
to consider the order parameter in the band basis because the 
superconductivity is mainly induced by the intra-band Cooper pairing 
when $|\Delta| \ll |\alpha|$. 
 Ignoring the order parameters describing the inter-band pairing, 
we obtain the simplified Hamiltonian as, 
\begin{eqnarray}
\label{eq:band-base}
&& \hspace*{-5mm} 
H_{\rm band} = \sum_{\gamma=1}^{4} \sum_{\k}{'} \ee_{\gamma}(\k) 
a_{\gamma,\k}^{\dag}a_{\gamma,\k} 
\nonumber \\ && \hspace*{7mm} 
 - [\Delta_{\gamma}(\k) a_{\gamma,-\km}^{\dag}a_{\gamma,\kp}^{\dag}
+ h.c.],   
\end{eqnarray}
where $\sum_{\k}{'}$ is restricted to the summation within 
$|k_{\rm z}| < \pi/2$. 
The dispersion relation $\ee_{\gamma}(\k)$ takes into account the ASOC and 
AFM order and is obtained by the unitary transformation as,  
\begin{eqnarray}
\label{eq:4-unitary}
\hat{U}^{\dag}(\k) \hat{H}(\k) \hat{U}(\k) = (\ee_{i}(\k) \delta_{ij}),  
\end{eqnarray}
where the $4 \times 4$ matrix $ \hat{H}(\k) $ is expressed as, 
\begin{eqnarray}
\label{eq:H-matrix}
\hat{H}(\k)
= 
\left(
\begin{array}{cc}
\hat{e}(\k) &
-\vec{h}_{\rm Q} \vec{\sigma} \\
-\vec{h}_{\rm Q} \vec{\sigma} & 
\hat{e}(\k+\Q) \\
\end{array}
\right). 
\end{eqnarray} 
We define $\hat{e}(\k)=\e(\k)\hat{\sigma}^{(0)} 
+ \alpha \vec{g}(\k) \vec{\sigma}$ 
and $\vec{\sigma}$ represent the three Pauli matrices and $\hat{\sigma}^{(0)} $ is the $ 2 \times 2 $-unit matrix. 
The four bands $\ee_{\gamma}(\k)$ are non-degenerate except for the 
special momentum, if $\alpha \ne 0$.  
Moreover, the relation $\ee_{\gamma}(-\k)=\ee_{\gamma}(\k)$ is hold owing to the 
time-reversal symmetry.

 The order parameter is expressed in the band basis as, 
\begin{eqnarray}
\label{eq:Delta-unitary}
\hat{\Delta}_{\rm band}(\k) 
= 
\hat{U}^{\dag}(\kp) \hat{\Delta}_{\rm spin}(\k) \hat{U}^{*}(-\km),
\end{eqnarray}
where
\begin{eqnarray}
\label{eq:Delta-matrix}
\hat{\Delta}_{\rm spin}(\k) 
= 
\left(
\begin{array}{cc}
\Delta_{1,s,s'}(\k) & \Delta_{2,s,s'}(\k) \\
\Delta_{2,s,s'}(\k+\Q) & \Delta_{1,s,s'}(\k+\Q) \\
\end{array}
\right).
\end{eqnarray}
 Although the off-diagonal matrix element of $\hat{\Delta}_{\rm band}(\k)$ 
is finite in general, the low-energy properties below \Tc are hardly 
affected by the off-diagonal components when $|\Delta| \ll |\alpha|$. 
 Therefore, we simply drop the off-diagonal components and obtain 
the Hamiltonian eq.~(6). 
 The SC order parameter for each band is expressed by 
the diagonal components as,   
$\Delta_{\gamma}(\k)=(\hat{\Delta}_{\rm band}(\k))_{\gamma\gamma}$.

 The normal and anomalous Green functions are expressed in the band basis as,
\begin{eqnarray}
\label{eq:Green-band}
&& \hspace*{-8mm}
G_{\gamma}(\kp,{\rm i}\omega_{n})=
({\rm i} \omega_{n} + \ee_{\gamma}(-\km))/A_{\gamma}(\k,{\rm i}\omega_{n}),
\\
&&\hspace*{-8mm}
F_{\gamma}(\k,{\rm i}\omega_{n})=
-\Delta_{\gamma}(\k)/A_{\gamma}(\k,{\rm i}\omega_{n}),
\end{eqnarray}
with 
\begin{eqnarray}
&&  \hspace*{-8mm}
A_{\gamma}(\k,{\rm i}\omega_{n})=
({\rm i} \omega_{n} - \ee_{\gamma}(\kp))({\rm i} \omega_{n} 
+ \ee_{\gamma}(-\km))
-|\Delta_{\gamma}(\k)|^{2}, 
\nonumber \\
\end{eqnarray}
where $\omega_{n}=(2 n + 1) \pi T$ is 
the Matsubara frequency and $T$ is the temperature.

 We decompose the uniform spin susceptibility 
into the Pauli part and Van-Vleck part, 
\begin{eqnarray}
\label{eq:uniform-susceptibility}
&&
\chi_{\mu\nu}=\chi_{\mu\nu}^{\rm P} + \chi_{\mu\nu}^{\rm V}. 
\end{eqnarray} 
 The Pauli susceptibility $\chi_{\mu\nu}^{\rm P}$ 
arises from the intra-band scattering while 
the inter-band scattering gives rise to the Van-Vleck susceptibility (VVS) 
$\chi_{\mu\nu}^{\rm V}$. 
 In the following we assume the staggered moments along the principal axis, 
namely $\vec{h}_{\rm Q} \parallel \hat{x}, \hat{y}$ or $\hat{z}$. 
 Following the Appendix A, the Pauli susceptibility and VVS 
are expressed as, 
\begin{eqnarray}
\label{eq:Pauli-part}
&&\hspace*{-10mm}
\chi_{\mu\mu}^{\rm P}=
-{\lim}_{\q \rightarrow 0}
{\lim}_{\Omega_{n} \rightarrow 0}
\sum_{\gamma,\k}{'}
A^{\mu\mu}_{\gamma\gamma}(\k)
\nonumber \\ && \hspace*{2mm} \times
[G_{\gamma}(k+q)G_{\gamma}(k) \pm 
F_{\gamma}(k+q)F^{\dag}_{\gamma}(k)], 
\end{eqnarray}
and 
\begin{eqnarray}
\label{eq:Van-Vleck-part}
&& \hspace{-12mm}
\chi_{\mu\nu}^{\rm V}= 
\sum_{\gamma \ne \delta} \sum_{\k}{'} 
A^{\mu\nu}_{\gamma\delta}(\k)
\frac{f(e_{\gamma}(\k))-f(e_{\delta}(\k))}{e_{\delta}(\k)-e_{\gamma}(\k)}, 
\end{eqnarray}
respectively. 
 The sign in eq.(15) is $+$ for $\mu=x,y$ and $-$ for $\mu=z$. 
 We define $A^{\mu\nu}_{\gamma\delta}(\k)=
S^{\mu}_{\gamma\delta}(\k,\k) S^{\nu}_{\delta\gamma}(\k,\k)$ where 
$S^{\mu}_{\gamma\delta}(\k+\q,\k)$ is the spin operator 
in the band basis, 
\begin{eqnarray}
\label{eq:spin-operator-band}
&&
\hat{S}^{\mu}(\k+\q,\k)=\hat{U}^{\dag}(\k+\q) 
\hat{S}^{\mu}
\hat{U}(\k), 
\end{eqnarray}
with 
\begin{eqnarray}
\hat{S}^{\mu}=
\left(
\begin{array}{cc}
\hat{\sigma}^{(\mu)} & 0 \\
0 & \hat{\sigma}^{(\mu)} \\
\end{array}
\right). 
\end{eqnarray}

 The expression of Pauli susceptibility eq.~(15) is equivalent to 
the spin susceptibility in a multi-band system. 
 The sign $+$ and $-$ in eq.~(15) correspond to the centrosymmetric 
superconductor with spin singlet pairing and that with spin triplet pairing 
for $\vec{d} \perp \vec{H}$, respectively. 
 Thus, the Pauli susceptibility of NCSC decreases 
in the {\it ab}-plane below \Tc while that is constant for the magnetic 
field along the {\it c}-axis.

 It should be noted that the VVS has a temperature dependence above \Tc 
which is similar to that of Pauli susceptibility, 
because the ASOC is much smaller than 
the Fermi energy ($|\alpha| \ll \e_{\rm F}$).~\cite{rf:fujimotoChi} 
 Therefore, the VVS in eq.~(14) should be included in the 
spin part of magnetic susceptibility which is extracted by the $K$-$\chi$ 
plot. 
 In this sense, the VVS, arising from the band splitting 
due to the ASOC, is quite different from the better known VVS 
coming from the orbital degrees of freedom. 
 Note that both VVS are not affected by 
the superconductivity when $T_{\rm c} \ll |\alpha|$.

 If the order parameter is spatially uniform, namely $\qfflo =0$, 
the Pauli susceptibility is described by the momentum dependent 
Yosida function as, 
\begin{eqnarray}
\label{eq:pauli-Yosida-xx}
&& \hspace*{-10mm}
\chi_{\mu\mu}^{\rm P}=
\sum_{\gamma} \int {\rm d}\k_{\rm F} 
A^{\mu\mu}_{\gamma\gamma}(\k_{\rm F})
Y(\Delta_{\gamma}(\k_{\rm F}),T)/v_{\gamma}(\k_{\rm F}), 
\end{eqnarray}
for $\mu = {\rm x,y}$, and 
\begin{eqnarray}
\label{eq:pauli-Yosida-zz}
&&\hspace*{-25mm}
\chi_{\rm zz}^{\rm P}=
\sum_{\gamma} \int {\rm d}\k_{\rm F} 
A^{\rm zz}_{\gamma\gamma}(\k_{\rm F})
/v_{\gamma}(\k_{\rm F}),
\end{eqnarray}
where $\int {\rm d}\k_{\rm F}$ is the integral on the Fermi surface, 
and $v_{\gamma}(\k_{\rm F})$ is the Fermi velocity of $\gamma$-th band. 
The Yosida function is defined as, 
\begin{eqnarray}
\label{eq:Yosida-function}
&&\hspace*{-15mm}
Y(\Delta,T)=-\int {\rm d}\e f'(\sqrt{\e^{2}+\Delta^{2}}), 
\end{eqnarray}
where $f'(E) = df/dE$ is the derivative of 
the Fermi distribution function. 
 Since $Y(\Delta,0)=0$ and $Y(0,T)=1$, we obtain 
$\chi_{\mu\mu}(T=0)=\chi_{\mu\mu}^{\rm V}$ for $\mu = x,y$ and 
$\chi_{\rm zz}(T=0)=\chi_{\rm zz}^{\rm V}
+\chi_{\rm zz}^{\rm P}(T=T_{\rm c})=\chi_{\rm zz}(T=T_{\rm c})$. 
 Thus, the residual spin susceptibility along {\it ab}-plane is 
given by the VVS alone, while for fields parallel to the {\it c}-axis 
both the Pauli and Van-Vleck susceptibility contribute. 
 It should be noticed that the spin susceptibility at $T=0$ is 
independent of the pairing symmetry. 
 In this sense the spin susceptibility is {\it universal} in the linear 
response regime when the system lacks the inversion 
symmetry.~\cite{rf:comment}

\subsection{PM state}

We concentrate now on the uniform state 
($\qfflo=0$) to investigate the residual spin susceptibility 
$\chi_{\mu\mu}$ at $T=0$ for $\mu = x,y$. 
The helical SC state with $\qfflo \ne 0$ will be discussed later in \S8. 
In the PM state we set $\vec{h}_{\rm Q}=0$ and assign the four bands as 
$\ee_{1,2}(\k)=\e(\k) \pm \alpha |\vec{g}(\k)|$, 
$\ee_{3,4}(\k)=\e(\k+\Q) \pm \alpha |\vec{g}(\k+\Q)|$ so that we can 
express the unitary matrix as,
\begin{eqnarray}
\label{eq:unitary-paramagnetic}
&&
\hat{U}(\k) =
\left(
\begin{array}{cc}
\hat{U}_{2}(\k) & 0 \\
0 & \hat{U}_{2}(\k+\Q) \\ 
\end{array}
\right),
\end{eqnarray}
where 
\begin{eqnarray}
\label{eq:2-unitary}
&& \hspace*{-14mm}
\hat{U}_{2}(\k) = \frac{1}{\sqrt{2}}
\left(
\begin{array}{cc}
1 & 1\\
\tilde{g}_{\rm x}(\k)+{\rm i} \tilde{g}_{\rm y}(\k) &  
-\tilde{g}_{\rm x}(\k)-{\rm i} \tilde{g}_{\rm y}(\k)\\ 
\end{array}
\right),  
\end{eqnarray}
with $\tilde{g}_{\mu}(\k)=g_{\mu}(\k)/|\vec{g}(\k)|$. 
 The matrix element of spin operator is obtained as,
$S^{\mu}_{11}(\k,\k)=-S^{\mu}_{22}(\k,\k) 
=\tilde{g}_{\mu}(\k)$ 
and 
$S^{\mu}_{33}(\k,\k)=-S^{\mu}_{44}(\k,\k)
=\tilde{g}_{\mu}(\k+\Q)$ for $\mu=x, y$ while 
$S^{\rm z}_{\gamma\gamma}(\k,\k) = 0$. 
We then find 
$\chi_{\rm xx}^{\rm P}=\chi_{\rm yy}^{\rm P} = \rho/2$ at $T=T_{\rm c}$ 
where $\rho$ is the DOS in the normal state. 
 Since $\chi_{\rm xx}=\chi_{\rm xx}^{\rm P}+\chi_{\rm xx}^{\rm V} = 
\rho + O(\alpha^{2}/\varepsilon_{\rm F}^{2})$ in the normal state, 
the residual spin susceptibility at $T=0$ is obtained as, 
\begin{eqnarray}
\label{eq:para-susceptibility}
\chi_{\rm xx}(T=0)=\chi_{\rm xx}^{\rm V}=\chi_{\rm xx}(T=T_{\rm c})/2 
+ O(\alpha^{2}/\varepsilon_{\rm F}^{2}). 
\end{eqnarray}
 Thus, the spin susceptibility in the {\it ab}-plane at $T=0$ 
is half of the normal state value in the limit 
$|\alpha| \ll \varepsilon_{\rm F}$. 
 Qualitatively the same result has been obtained in Refs.~18,20,22-28. 
 Fujimoto has shown that the VVS increases when the 
DOS has strong asymmetry and $|\alpha|$ is moderate.~\cite{rf:fujimotoChi} 
However, the $\beta$-band of CePt$_3$Si which we will investigate later 
does not satisfy this condition.~\cite{rf:yanaselett}

 Since the spin susceptibility decreases below \Tc for the magnetic field 
along the {\it ab}-plane, the paramagnetic depairing effect of 
$H_{\rm c2}$ should be observed in NCSC 
with Rashba-type spin-orbit coupling. 
 This is consistent with the recent observation of the 
paramagnetic depairing effect in CeRhSi$_3$~\cite{rf:kimurareview,
rf:kimuraprivate} and 
CeIrSi$_3$~\cite{rf:okuda,rf:onukireview,rf:onukiprivate} 
under high pressure where the AFM order is suppressed. 
 However, this is not the case in CePt$_3$Si at ambient pressure 
(within the AFM phase).~\cite{rf:yogiK,rf:higemoto,rf:bauerDC,
rf:bauerreview,rf:onukireview,rf:yasuda} 
 This observation leads us to study the influence of AFM order 
in the following part.

\subsection{AFM state with inversion symmetry}

 In order to clarify the influence of AFM order, we first investigate the 
spin susceptibility in the SC state {\it with} inversion symmetry for 
$ \vec{h}_{\rm Q} \neq 0 $.
 Owing to the inversion symmetry, the residual spin susceptibility depends 
on the pairing symmetry in the usual way. 
 Here we discuss the spin singlet pairing state while the spin 
susceptibility is constant through \Tc in the spin triplet pairing state. 
 The spin susceptibility consists of the Pauli part and 
Van-Vleck part as in \S2.1, and the Pauli part vanishes in the spin 
singlet pairing state at $T=0$.

 As a result of the simple calculation, we obtain for 
the Pauli susceptibility above \Tcf, 
\begin{eqnarray}
\chi_{\mu\mu}^{\rm P}= \sum_{\k} [\delta(e_{1}(\k))+\delta(e_{2}(\k))] 
= \rho, 
\end{eqnarray}
for $\vec{H} \parallel \vec{h}_{\rm Q}$ 
and 
\begin{eqnarray}
\chi_{\mu\mu}^{\rm P}=\sum_{\k} 
\frac{\varepsilon_{-}(\k)^{2}}{\varepsilon_{-}(\k)^{2}+h_{\rm Q}^{2}}
[\delta(e_{1}(\k))+\delta(e_{2}(\k))], 
\end{eqnarray}
for $\vec{H} \perp \vec{h}_{\rm Q}$. 
 Here, $e_{1,2}(\k)=\varepsilon_{+}(\k) \pm \sqrt{\varepsilon_{-}(\k)^{2}+h_{\rm Q}^{2}}$ with 
$h_{\rm Q}=|\vec{h}_{\rm Q}|$ and 
$\varepsilon_{\pm}(\k) = (\varepsilon(\k) \pm \varepsilon(\k+\Q))/2$. 
The Pauli part of spin susceptibility 
for the magnetic field perpendicular to the AFM moment  decreases
with growing $ \vec{h}_{\rm Q}$, i.e., 
 $\chi_{\mu\mu}^{\rm P}(h_{\rm Q}=0) > 
\chi_{\mu\mu}^{\rm P}(h_{\rm Q} \ne 0)$ for 
$\vec{H} \perp \vec{h}_{\rm Q}$. 
A Van-Vleck part is induced by the AFM order 
and leads to the residual spin susceptibility 
for $\vec{H} \perp \vec{h}_{\rm Q}$ at $T=0$. 
In contrast, the Van-Vleck part and the residual spin susceptibility 
vanish for the magnetic field parallel to the AFM moment.

In Fig.~1 we show the numerical results for the spin susceptibility in the 
SC state at $T=0$. For this numerical analysis 
we assume a tight-binding model approximating the so-called $\beta$-band 
of CePt$_3$Si,~\cite{rf:yanaselett}  
\begin{eqnarray}
\label{eq:dispersion}
&& \hspace*{-8mm}  \e(\k)  =   2 t_1 (\cos k_{\rm x} +\cos k_{\rm y}) 
         + 4 t_2 \cos k_{\rm x} \cos k_{\rm y} 
\nonumber \\ && \hspace*{-8mm} 
         + 2 t_3 (\cos 2 k_{\rm x} +\cos 2 k_{\rm y})
+ [ 2 t_4 + 4 t_5 (\cos k_{\rm x} +\cos k_{\rm y}) 
\nonumber \\ && \hspace*{-8mm} 
         + 4 t_6 (\cos 2 k_{\rm x} +\cos 2 k_{\rm y}) ] \cos k_{\rm z}
         + 2 t_7 \cos 2 k_{\rm z} 
         - \mu_{\rm c}. 
\end{eqnarray}
 The chemical potential $\mu_{\rm c}$ is determined so that the electron 
density per site is $n$. 
 The Fermi surface of the $\beta$-band, which has been obtained in 
the band structure calculations without taking AFM order into 
account,~\cite{rf:samokhinband,rf:anisimov,rf:hashimotoDHV} 
is reproduced by choosing the parameters as 
\begin{eqnarray}
&& \hspace*{-15mm} (t_1,t_2,t_3,t_4,t_5,t_6,t_7,n) = 
\nonumber \\ && \hspace*{-15mm}
(1,-0.15,-0.5,-0.3,-0.1,-0.09,-0.2,1.75), 
\end{eqnarray}
and $\alpha=0.3$ 
defining $ t_1 $ as the unit energy. 
 As shown in Fig.~1, the spin susceptibility for the magnetic field 
perpendicular to the AFM moment 
$m = \chi_{\rm xx}(\vec{Q},0) h_{\rm Q}^{\rm x}$ is 
increased in the SC state 
by the AFM order while that in the normal state is little affected.

\begin{figure}[ht]
\begin{center}
\includegraphics[width=7cm]{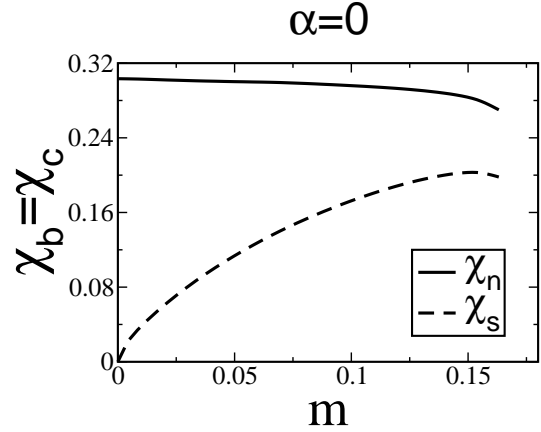}
\caption{
 Spin susceptibility along the {\it b}- and {\it c}-axes at $T=0$ against the 
staggered magnetic moment $m$ along the {\it a}-axis. 
 The ASOC is zero ($\alpha=0$) and the staggered field 
$\vec{h}_{\rm Q} = h_{\rm Q}^{\rm x}\hat{x}$ is assumed. 
 The staggered magnetic moment is defined as 
$m=|<\sum_{s,s'}\sigma^{\rm (x)}_{ss'} c_{i,s}^{\dag}c_{i,s'} >|$ 
so that $m=1$ is the full moment. 
 The solid and dashed lines show the spin susceptibility in the 
normal state and in the spin singlet SC state, respectively. 
 The spin susceptibility along the {\it a}-axis is zero in the spin 
singlet SC state at $T=0$. 
} 
\end{center}
\end{figure}

At this point we can discuss the role of the band structure. 
 According to eq.~(26), the Pauli part of the spin susceptibility is small for the magnetic field
$\vec{H} \perp \vec{h}_{\rm Q}$, if 
the quasiparticle dispersion is quasi-two dimensional 
and $\varepsilon_{-}(\k)$ is small. 
 Although the $\beta$-band of CePt$_3$Si has a three dimensional Fermi 
surface, the band dispersion is weak along the $k_{\rm z}$-axis 
according to the result of band calculation.~\cite{rf:samokhinband}
 This means that the AFM order significantly would affect the SC  
state in this band of CePt$_3$Si. 
 The presence of quasi-two dimensional Fermi surface is also 
expected in CeRhSi$_3$.~\cite{rf:harimaprivate}

There are cases where AFM order plays indeed an important role 
in a centrosymmetric material. 
For example, UPd$_2$Al$_3$ is a spin singlet superconductor with 
$T_{\rm c}=2$K which coexists with the AFM order.~\cite{rf:Geibel} 
 The AFM state has a high N\'eel temperature of
$T_{\rm N}=14$K and a large staggered magnetic moment, 
$m=0.85 \mu_{\rm B}$. 
 This moment is directed to the {\it ab}-plane of 
tetragonal lattice and $\Q = (0,0,\pi)$.~\cite{rf:Krimmel} 
 This is the same spin structure as CePt$_3$Si. 
NMR measurements show the decrease of Knight shift below
\Tc with a large residual part.~\cite{rf:tou} 
 The VVS arising from the AFM order may induce 
the large residual spin susceptibility, 
although the multi-orbital effect is another possible origin. 
 This is consistent with the large $H_{\rm c2}$ which exceeds 
the standard paramagnetic limit.~\cite{rf:amato}

\subsection{AFM state without inversion symmetry}

 The result in \S2.3 implies that the AFM order enhances the 
Van-Vleck part of spin susceptibility in the non-centrosymmetric 
system for the magnetic field perpendicular to the AFM moment. 
 We have shown the results for the spin susceptibility 
along the {\it a}- and {\it b}-axes by assuming the dispersion relation 
eqs.~(27), (28), $\alpha=0.3$ and $\vec{h}_{\rm Q} \parallel \hat{x}$ 
to describe the electronic structure of CePt$_3$Si below $T_{\rm N}$ 
(Fig.~4 of Ref.~27). 
 For fields $\vec{H} \perp \vec{h}_{\rm Q}$ the normal state and SC state 
susceptibility merge for increasing staggered moment, suggesting 
a diminishing of the reduction of the spin susceptibility in the SC state. 
 On the other hand, the behavior is opposite for 
$\vec{H} \parallel \vec{h}_{\rm Q}$. 
 Thus, a remarkable 2-fold anisotropy is expected in the spin 
susceptibility below \Tc even if the anisotropy is weak in 
the normal state. 
 The condition $\vec{H} \perp \vec{h}_{\rm Q}$ is generally favored because 
the magnetization energy is maximally gained for the field direction 
with largest spin susceptibility. 
 However, the meta-stable state $\vec{H} \parallel \vec{h}_{\rm Q}$ can be 
realized in the weak magnetic field which is smaller than 
the anisotropy energy of AFM moment.

 The role of AFM order is suppressed by increasing the ASOC. 
 We have confirmed that $\chi_{\rm yy}$ in the SC state is decreased 
by increasing $\alpha$ when $\vec{h}_{\rm Q} \parallel \hat{x}$. 
 The AFM order plays a quantitatively important role when the ASOC 
is much smaller than the Fermi energy. 

If the AFM moment is parallel to the {\it c}-axis as in 
CeCoGe$_3$,~\cite{rf:CeCoGe} 
the spin susceptibility along both {\it a}- and {\it b}-axes 
is increased in the SC state by the AFM order, while that along the 
{\it c}-axis is not affected.

\section{Effective Model for CePt$_3$Si, CeRhSi$_3$ and CeIrSi$_3$}

In preparation for the discussion of the non-linear response to the magnetic field 
we will introduce here an effective model for
 CePt$_3$Si, CeRhSi$_3$ and CeIrSi$_3$. 
This is important as we will show that the non-linear spin susceptibility 
significantly depends 
on the symmetry of order parameter in contrast to the {\it universal} 
spin susceptibility in the linear response theory (see \S2.1). 
 In particular, we point out the strong non-linearity in the pairing state with dominant
$p$-wave character.

 We analyze the following effective model, 
\begin{eqnarray}
\label{eq:Hubbard-model}
&& \hspace*{-5mm}  
H = H_{\rm b} + H_{\rm SO} + H_{\rm AF} + H_{\rm Z} + H_{I},
\\ && \hspace*{-5mm} 
H_{\rm Z} = - \sum_{\vec{k}} \h  \cdot \vec{S}(\k),
\\ && \hspace*{-5mm} 
H_{\rm I} = 
    U \sum_{i} n_{i,\uparrow} n_{i,\downarrow}
   + (V-J/4) \sum_{<i,j>} n_{i} n_{j} 
\nonumber \\ && \hspace*{3mm} 
   + J \sum_{<i,j>} (\vec{S}_{i} \cdot \vec{S}_{j}
   - 2 S^{\rm x}_{i} S^{\rm x}_{j}) 
\\ && \hspace*{1mm} 
= \frac{1}{2} \sum_{\k,\k',\q,s} 
[V C(\k-\kk)
c_{-\km,s}^{\dag} c_{\kp,s}^{\dag} 
c_{\kpk,s} c_{-\kmk,s} 
\nonumber \\ && \hspace*{3mm} 
+
\{U + (V - \frac{J}{2}) C(\k-\kk)\}
c_{-\km,\bar{s}}^{\dag} c_{\kp,s}^{\dag} 
c_{\kpk,s} c_{-\kmk,\bar{s}} 
\nonumber \\ && \hspace*{3mm} 
-
\frac{J}{2}  C(\k-\kk) 
c_{-\km,s}^{\dag} c_{\kp,s}^{\dag} 
c_{\kpk,\bar{s}} c_{-\kmk,\bar{s}}], 
\end{eqnarray}
where $n_{i,s}$ is the electron number at the site $i$ with spin $s$, 
$n_{i}=n_{i,\uparrow}+n_{i,\downarrow}$, $\bar{s}=-s$ and 
$C(\k)=2 (\cos k_{\rm x} +  \cos k_{\rm y})$. 
 The spin operator in the real space basis is defined as 
$\vec{S}_{i}= \sum_{s,s'} \vec{\sigma}_{ss'} c_{i,s}^{\dag}c_{i,s'}$.  
 The bracket $<i,j>$ denotes the summation for the nearest neighbor sites  
in the {\it ab}-plane, namely $j=i \pm \vec{a}$ or $j=i \pm \vec{b}$ with 
$\vec{a}$ and $\vec{b}$ the unit vectors along the {\it a}- and 
{\it b}-axis, respectively.

 The first three terms in eq.~(29) have been defined earlier in eqs.~(2-4). 
 For the dispersion relation, we adopt the tight-binding model eq.~(27) 
with using the parameter set eq.~(28) and $\alpha=0.3$, 
reproducing the $\beta$-band of 
CePt$_3$Si.~\cite{rf:samokhinband,rf:hashimotoDHV,rf:anisimov} 
 We choose the $\beta$-band, because it has substantial Ce 
4$f$-electron character~\cite{rf:anisimov} and the largest DOS at the 
Fermi energy, namely 70\% of the total DOS.~\cite{rf:samokhinband} 
Besides the sizable jump in specific heat, also 
the remarkable isotropy of $H_{\rm c2}$ between the {\it ab}-plane 
and {\it c}-axis~\cite{rf:yasuda} also indicates that the 
three-dimensional Fermi surface of the $\beta$-band is mainly responsible 
for the superconductivity in \Ptf. 
 In Appendix B we will investigate the other dispersion relation 
which favors the $d_{\rm x^{2}-y^{2}}$-wave superconductivity.

 As for the AFM order, we assume the staggered field pointing along the 
[100]-direction $\vec{h}_{\rm Q}=h_{\rm Q}\hat{x}$ with 
 $ \Q = (0,0, \pi) $ following the experimental 
results of CePt$_3$Si.~\cite{rf:metoki} 
 For the magnitude we assume $h_{\rm Q} \ll W $,  the band 
width. This is consistent with the small observed magnetic moment 
$\sim 0.16 \mu_{\rm B}$ in \Ptf.~\cite{rf:metoki} 
 The AFM moment is expected to be small also in \Rh and \Ir 
since superconductivity occurs near the AFM quantum critical point. 

 The fourth term $H_{\rm Z}$ is the Zeeman coupling term due to the 
applied magnetic field. We have defined 
$\h = \frac{1}{2} g \mu_{\rm B} \vec{H}$ where $g$ is the $g$-factor of 
quasiparticles and $\mu_{\rm B}$ is the Bohr magneton. 
 The paramagnetic depairing effect on the superconductivity is 
characterized by the dimensionless 
coupling constant $h/T_{\rm c}$ with $h=|\vec{h}|$.

 The last term $H_{\rm I}$ describes the effective interaction leading 
to the SC instability and includes three coupling constants, 
$U$, $V$ and $J$. 
 We assume the on-site interaction $U$ and the interaction between the 
nearest neighbor sites in the {\it ab}-plane $V$. 
 The coupling constant $J$  describes the part of 
interaction arising from the AFM order which is anisotropic. 
 According to the random phase approximation (RPA) 
for the Hubbard model,~\cite{rf:yanaselett,
rf:yanasefull} the SC order parameter is affected by the AFM order mainly 
through the anisotropy of effective interaction, which can be described by 
the $J$-term in eq.~(31).

In the following we examine two parameter sets, 
\begin{eqnarray}
&&\hspace*{-20mm} {\rm (A)}  
\hspace*{3mm}U>0,  \hspace*{2mm}V=-0.8 U, 
\\
&&\hspace*{-20mm} {\rm (B)} 
\hspace*{3mm}U<0, \hspace*{2mm}V=0.
\end{eqnarray}
 The amplitude of $U$ is chosen so that $T_{\rm c}=0.01$ 
at zero magnetic field. 
 The ground state is dominantly (A) $p$-wave and (B) $s$-wave, respectively. 
 Hereafter we simply call these states $p$-wave and $s$-wave 
state, respectively. 
 The parameter set (A) is the most important for our purpose, because 
the $p$-wave symmetry is the most promising candidate for the pairing state in 
\Ptf.~\cite{rf:hayashiT1,rf:fujimoto,rf:sigristpro,rf:fujimotoreview,
rf:yanaselett} 
 Although the spin triplet superconductivity is handicapped due to the lack of 
inversion symmetry in non-centrosymmetric systems according to the 
Anderson's theorem,~\cite{rf:anderson} the depairing effect arising 
from the ASOC vanishes (or is at least smallest) in the $p$-wave state with 
$\vec{d}(\k) \parallel \vec{g}(\k)$.~\cite{rf:frigeri} 
 This condition is not satisfied in the realistic model, 
however the depairing effect due to the ASOC is almost avoided in the 
$p$-wave state with 
$\vec{d}(\k) = - p_{\rm y}\hat{x} + p_{\rm x}\hat{y}$.~\cite{rf:yanaselett}
 Another parameter set (B) is investigated as a typical 
model for the dominantly spin singlet pairing state. 
 We will investigate the dominantly $d_{\rm x^{2}-y^{2}}$-wave state in 
Appendix B and obtain qualitatively the same results as the $s$-wave 
state.

 Before analyzing the effective model in eq.~(29), we comment on the 
RPA theory applied to the Hubbard model for the $\beta$-band 
of CePt$_3$Si.~\cite{rf:yanaselett,rf:yanasefull} 
This theory leads to two possible pairing states due to spin fluctuation mediated
interaction: the $s+P$-wave 
 and the $p+D+f$-wave state. The former is dominated by the $p$-wave component and 
can be viewed as an intra-plane pairing state, while the 
latter is described by the inter-plane pairing dominated by the 
$d_{\rm xz}$- and $d_{\rm yz}$-wave components. 
Here we focus on the $s+P$-wave state whose order parameter is 
reproduced by assuming the parameter set (A) and $J=0.3 V$ ($J=0$) for 
$h_{\rm Q}=0.125$ ($h_{\rm Q}=0$) in eq.~(29). 
 On the other hand, the $p+D+f$-wave state is not realized
in eq.~(29) because the inter-plane interaction is neglected. 
 It is expected that the paramagnetic properties in the 
inter-plane $d$-wave state are qualitatively the same as those 
in the intra-plane $s$- and $d$-wave states. 
 The other characteristic properties of the $p+D+f$-wave state will be 
discussed in \S6 and \S7.

 To solve the effective model eq.~(29), we apply the mean field theory 
and obtain the mean field equations as, 
\begin{eqnarray}
\label{eq:mf-equation}
&& \hspace*{-8mm}  
\Delta_{i,s,\bar{s}}(\k) = 
- T \sum_{n,\kk} 
\{U + (V - \frac{J}{2}) C(\k-\kk)\} 
\nonumber \\ && \hspace*{20mm} \times
F_{i,s,\bar{s}}(\kk,\omega_{n}),
\\ && \hspace*{-8mm} 
\Delta_{i,s,s}(\k) = 
- T \sum_{n,\kk} C(\k-\kk) \{V F_{i,s,s}(\kk,\omega_{n}) 
\nonumber \\ && \hspace*{20mm} 
-\frac{J}{2} F_{i,\bar{s},\bar{s}}(\kk,\omega_{n})\}.  
\end{eqnarray}
The normal and anomalous Green functions 
$G_{i,s,s'}(\kk,\omega_{n})$,  
$F_{i,s,s'}(\kk,\omega_{n})$ in the spin basis are 
obtained by the Dyson-Gorkov equation,
\begin{eqnarray}
\label{eq:dg-equation}
&& \hspace*{-5mm}  
\left(
\begin{array}{cc}
\hat{G}_{\rm N}(\kp,\omega_{n})^{-1} & \hat{\Delta}_{\rm spin}(\k) \\
\hat{\Delta}_{\rm spin}^{\dag}(\k)  
& -\hat{G}_{\rm N}^{\rm T}(-\km,-\omega_{n})^{-1} \\
\end{array}
\right) 
\nonumber \\ &&  \hspace*{-5mm}
\times
\left(
\begin{array}{cc}
\hat{G}(\kp,\omega_{n}) & \hat{F}(\k,\omega_{n}) \\
\hat{F}^{\dag}(\k,\omega_{n})  & -\hat{G}^{\rm T}(-\km,-\omega_{n}) \\
\end{array}
\right) 
= 
\hat{1}. 
\end{eqnarray}
where $\hat{X}(\k)$ ($X=G,F,\Delta_{\rm spin}$) is the 4 $\times$ 4 matrix, 
\begin{eqnarray}
\label{eq:4by4-matrix}
\hat{X}(\k) 
= 
\left(
\begin{array}{cc}
X_{1,s,s'}(\k) & X_{2,s,s'}(\k) \\
X_{2,s,s'}(\k+\Q) & X_{1,s,s'}(\k+\Q) \\
\end{array}
\right). 
\end{eqnarray}
 The normal Green function in the normal state, 
$\hat{G}_{\rm N}(\k,\omega_{n})$ is obtained  as 
$\hat{G}_{\rm N}(\k,\omega_{n}) = 
({\rm i}\omega_{n} \hat{1} - \hat{H}(\k))^{-1} 
$
by using eq.~(8) with $\vec{h}_{\rm Q}=h_{\rm Q}\hat{x}$ and 
$\hat{e}(\k)=\e(\k)\hat{\sigma}^{(0)} 
+ \alpha \vec{g}(\k) \cdot \hat{\vec{\sigma}} - \vec{h} \cdot \hat{\vec{\sigma}}$.

 We here discuss the symmetry of the SC state on the basis of 
the following parameterization of order parameters: 
\begin{eqnarray}
\label{eq:d-vector}
\Delta_{1,s,s'}(\k)
= 
\left(
\begin{array}{cc}
-d_{{\rm x}}(\k)+{\rm i}d_{{\rm y}}(\k) & \Phi(\k) + d_{{\rm z}}(\k) \\
-\Phi(\k) + d_{{\rm z}}(\k) & d_{{\rm x}}(\k)+{\rm i}d_{{\rm y}}(\k) \\
\end{array}
\right), 
\end{eqnarray}
where we use the even parity scalar function $ \Phi (\k) $ 
and the odd parity vector $ \vec{d}(\k) $. 
 Although a second component $\Delta_{2,s,s'}(\k)$ appears in the AFM state, 
the basic properties and symmetries are little 
affected by $\Delta_{2,s,s'}(\k)$.

\begin{table}[htbp]
  \begin{center}
    \begin{tabular}{|c||c|c|} \hline 
P-wave state & Even parity part & Odd parity part  \\\hline\hline 
PM at $\vec{h}=0$ 
& $\kappa (\delta + {\rm c}_{\rm x}+{\rm c}_{\rm y}) $
& $(-{\rm s}_{\rm y},{\rm s}_{\rm x},0)$ \\\hline
AFM at $\vec{h}=0$ 
& $\kappa (\delta + \eta {\rm c}_{\rm x}+{\rm c}_{\rm y}) $
& $(-{\rm s}_{\rm y},\beta {\rm s}_{\rm x}, 0)$ \\\hline
AFM at $\vec{h} = h \hat{y}$ 
& $\kappa(\delta + \eta {\rm c}_{\rm x}+{\rm c}_{\rm y}) $
& $(-{\rm s}_{\rm y},\beta {\rm s}_{\rm x}, 
-{\rm i}\gamma {\rm s}_{\rm y})$ \\\hline
    \end{tabular}
\caption{
Symmetry of order parameter in the dominantly $p$-wave state. 
We use the abbreviations ${\rm c}_{\rm x,y} = \cos k_{\rm x,y}$ and 
${\rm s}_{\rm x,y} = \sin k_{\rm x,y}$.
We assume the PM state at zero magnetic field, the AFM state 
at zero magnetic field and the AFM state under the magnetic field 
$\vec{h} \parallel \hat{y}$ from the top to the bottom. 
The parameters $\beta$, $\gamma$, $\kappa$, $\delta$ and $\eta$ are real. 
}
  \end{center}
\end{table}

 In Table~I we summarize the order parameters in the $p$-wave state. 
 The admixture with the even-parity component due to the ASOC is expressed 
by the parameter $\kappa$ which is in the order of $\alpha/\e_{\rm F}$. 
 We obtain  $\kappa \sim 0.15$ for $\alpha=0.3$. 
 The even-parity part $ \Phi(\k) $ is dominated by the extended 
$s$-wave component and $\delta \sim 0.2$ 
because the conventional $s$-wave component is suppressed  by the 
strong on-site repulsion $U$. 
 The $d_{\rm z}$-component of the odd parity vector $\vec{d}(\k)$ 
is induced by the magnetic field $\gamma \propto h/\e_{\rm F}$ 
to gain the Zeeman energy. 
 The SC state is mainly affected by the parameter $\beta$ 
which is unity in the absence of AFM order and magnetic field. 
 The magnetic field along the [010]-axis ([100]-axis) decreases (increases) 
$\beta$. The influence of the AFM order depends on the value of $J$. 
 We find $\beta \sim 0.3$ ($\beta \sim 3.7$) for $J=0.3V$ ($J=0$) at 
$h_{\rm Q}=0.125$ and $h=0$. 
 The deviation from $\beta=1$ can be viewed as the mixing between 
$\vec{d}(\k) = (- \sin k_{\rm y},\sin k_{\rm x},0)$ and 
another $p$-wave state $\vec{d}(\k) = (\sin k_{\rm y},\sin k_{\rm x},0)$. 
Although these pairing states belong to different irreducible representations 
of the $D_{\rm 4h}$ symmetry, they are mixed due to the presence of the symmetry
reducing
AFM moment or magnetic field.

 In general, the dominantly $s$-wave state is admixed to the $p$-wave state 
due to the ASOC and belongs to the same irreducible representation as 
the dominantly $p$-wave state realized for the parameter set (A). 
 However, only the conventional $s$-wave component 
$\Phi(\k) = 1$ appears and $\vec{d}(\k)=\vec{0}$ for the parameter set (B) 
because only the on-site interaction is taken into account ($V=J=0$). 
 Generally speaking, the admixture of spin singlet and triplet order 
parameters plays no important role when the ASOC 
is much smaller than the Fermi energy, $|\alpha| \ll \e_{\rm F}$.

 The ''helicity'' $\qfflo$-vector is perpendicular to the magnetic field 
as will be discussed in \S8 in details. 
 The amplitude of $\qfflo$ below \Tc should be 
determined to maximize the condensation energy. 
 However, here we determine $\qfflo$ at $T=T_{\rm c}(h)$ and neglect the 
temperature dependence below \Tc for simplicity. 
 The transition temperature $T_{\rm c}(h)$ is determined by linearizing the 
mean field equation as,
\begin{eqnarray}
\label{eq:linearized-mf-equation}
&& \hspace*{-5mm}  
\lambda(\q) \Delta_{i,s,\bar{s}}(\k) = 
- T \sum_{n,\kk} 
\{U + (V - \frac{J}{2}) C(\k-\kk)\} 
\nonumber \\ && \hspace*{25mm} \times 
\phi_{i,s,\bar{s}}(\kk,\omega_{n}),
\\ && \hspace*{-5mm} 
\lambda(\q) \Delta_{i,s,s}(\k) = 
- T \sum_{n,\kk} C(\k-\kk) 
\nonumber \\ && \hspace*{10mm} \times 
\{V \phi_{i,s,s}(\kk,\omega_{n}) 
-\frac{J}{2} \phi_{i,\bar{s},\bar{s}}(\kk,\omega_{n})\}, 
\end{eqnarray}
where $\phi_{i,s,s'}(\kk,\omega_{n})$ is 
obtained by linearizing the anomalous Green function 
$F_{i,s,s'}(\kk,\omega_{n})$ with respect to $\hat{\Delta}_{\rm spin}(\kk)$. 
 We optimize the eigenvalue  $\lambda(\q)$ with respect to the 
order parameter $\hat{\Delta}_{\rm spin}(\k)$ and the helicity $\q =\qfflo$. 
 The transition temperature $T_{\rm c}(h)$ is determined by the criterion 
$\lambda(\qfflo)=1$.

 We have estimated the condensation energy below \Tc and found that 
the magnitude of $\qfflo$ increases as decreasing the temperature. 
 However, we have confirmed that the temperature dependence 
of $\qfflo$ can be ignored for the magnetic properties 
discussed in the following part.

\section{$S$-wave State}

 For the discussion of non-linear response to the magnetic field 
in NCSC we first discuss the simplest case, namely the $s$-wave 
state without AFM order. 
 We address the enhancement of the critical magnetic field 
$h_{\rm c2}=\frac{1}{2}g\mu_{\rm B}H_{\rm c2}$ due to the ASOC,
assuming the parameter set (B) $U < 0$ and $V=0$. 
 Figure~2 shows the phase diagram, temperature 
$T/T_{\rm c}$ versus magnetic
field $h/T_{\rm c}$ along the [100]- or [010]-direction.
The critical field \hc for both the uniform state ($\qfflo=0$) and
the helical state ($\qfflo \ne 0$) are depicted, whereby also the behavior 
in the absence of ASOC ($\alpha=0$) is included for a comparison. 
As we focus here on the paramagnetic limiting effect, we 
neglect the orbital depairing for simplicity. 
Note that  \hc in the helical $s$-wave state has been investigated 
in Ref.~35 for an isotropic Fermi surface.

\begin{figure}[htbp]
  \begin{center}
\includegraphics[width=6.5cm]{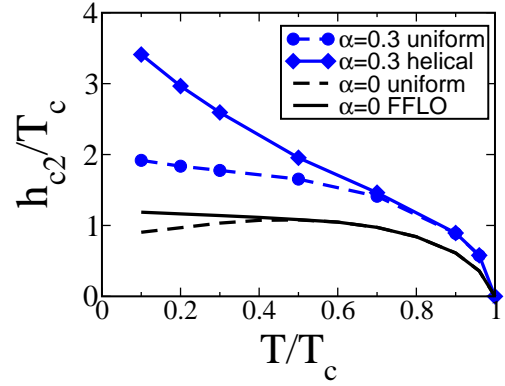}
\caption{(Color online)
 The $H$-$T$ phase diagram in the $s$-wave state for
the magnetic field along the [100]- or [010]-axis.
 The diamonds (circles) show the reduced critical magnetic field
$h_{\rm c2}/T_{\rm c}=\frac{1}{2}g \mu_{\rm B} H_{\rm c2}/T_{\rm c}$
in the helical (uniform) SC state against the reduced temperature
$T/T_{\rm c}$.
 We assume $U < 0$, $V=0$, $J=0$, $\alpha=0.3$ and $h_{\rm Q}=0$. 
 The phase diagrams in the absence of ASOC ($\alpha=0$) are shown 
for a comparison. 
The dashed and solid lines show the $h_{\rm c2}/T_{\rm c}$ 
in the uniform ($\qfflo=0$) and FFLO ($\qfflo \ne 0$) states, respectively. 
}
  \end{center}
\end{figure}

 The data in Fig.~2 demonstrate that the \hc is significantly enhanced 
by the ASOC. 
 This is partly due to the residual spin susceptibility in the SC state 
induced by the ASOC.  Neglecting the magnetic field dependence of the spin susceptibility, 
we obtain a simple estimation for the critical magnetic field, 
\begin{eqnarray}
h_{\rm c2} = \sqrt{\frac{2 E_{\rm c}}{\chi^{\rm N}-\chi^{\rm S}}}. 
\end{eqnarray}
where $E_{\rm c}$ is the condensation energy and 
$\chi^{\rm S}$ and $\chi^{\rm N}$ are 
the spin susceptibility in the SC and normal state, respectively. 
 According to eq.~(42),  \hc increases by a factor of $\sqrt{2}$ because of 
the residual spin susceptibility $\chi^{\rm S}=\frac{1}{2}\chi^{\rm N}$ 
at $T=0$.  In fact,  \hc is enhanced even more due to the magnetic field 
dependence of spin susceptibility. A further enhancement of \hc is caused 
by the formation of a helical SC state, which exceeds the enhancement in
centrosymmetric superconductor owing to the presence of an FFLO 
state.~\cite{rf:FFLO}

\begin{figure}[htbp]
  \begin{center}
\includegraphics[width=6.5cm]{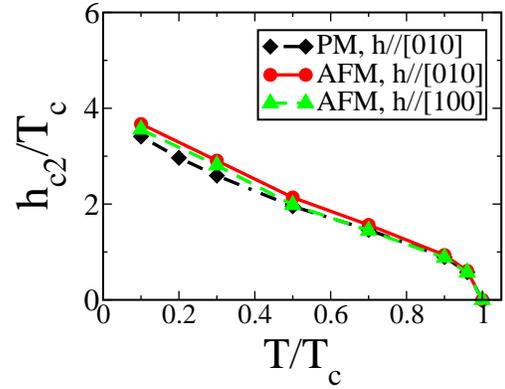}
\caption{(Color online)
The $H$-$T$ phase diagram in the helical $s$-wave state with AFM order. 
We assume $\vec{h}_{\rm Q}=0.125 \hat{x}$. 
 The other parameters are the same as Fig.~2. 
 The circles and triangles show the \hc for the magnetic field 
along the [010] and [100]-axis, respectively. 
 The \hc in the PM state is shown for a comparison (diamonds). 
}
  \end{center}
\end{figure}

 We here investigate the influence of AFM order on the $s$-wave SC state. 
 Figure~3 shows that the \hc in the $s$-wave state 
is increased by the AFM order, however the enhancement is very small. 
 According to the simple estimation eq.~(42) and the 
{\it universal} spin susceptibility in the linear response theory 
(see \S2.4), the \hc along the [010]-axis ([100]-axis) 
is enhanced (suppressed) by the AFM order 
through the increase (decrease) of $\chi^{\rm S}$. 
 However, the enhancement (suppression) is much smaller than expected 
in this simple estimation. 
 This is mainly because of the formation of helical SC phase 
which induces the non-linear spin susceptibility at high fields. 
 In \S5.3 we will show that the influence of AFM order is much more 
significant in the dominantly $p$-wave state.

\section{The $p$-wave State}

 We here investigate the dominantly $p$-wave state 
which is the most promising candidate for the 
pairing state in \Ptf. 
 In this section we assume the parameter set (A) $U > 0$, $V=-0.8 U$.

\subsection{PM state}

\begin{figure}[htbp]
  \begin{center}
\includegraphics[width=6.5cm]{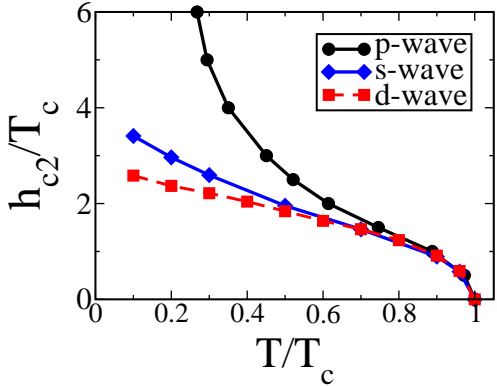}
\caption{(Color online)
 The $H$-$T$ phase diagram in the helical $p$-wave state (circles).
 We assume $U > 0$, $V=-0.8U$, $J=0$, $\alpha=0.3$ and $h_{\rm Q}=0$.  
 The \hc in the the $s$-wave state (diamonds) and in the 
$d_{\rm x^{2}-y^{2}}$-wave state (squares) are shown for a comparison. 
 The parameter set for the $d_{\rm x^{2}-y^{2}}$-wave state is shown 
in the Appendix B. 
}
  \end{center}
\end{figure}

 To illuminate the difference with the dominantly spin singlet pairing 
state we again turn to the PM state. 
 We find that paramagnetic depairing effect is naturally less effective 
in suppressing the onset of superconductivity. 
 Figure~4 shows  \hc for the $p$-wave state which is much higher than 
for the case of $s$-wave as well as $d$-wave pairings. 
 This is rather surprising because the \hc is independent of 
the pairing symmetry considering only the simple estimation in eq.~(42). 
 Actually,  \hc of the $p$-wave state is enhanced by 
the modification of SC order parameters due to the mixing with 
$\vec{d}(\k)=(\sin k_{\rm y}, \sin k_{\rm x},0)$ in addition to the 
formation of helical SC state. 
 Since the $p$-wave superconductivity has a multi-component
order parameter with respect to the spin, the order parameter 
can be modified to optimally cope with the competition 
between the Zeeman coupling energy and ASOC. 
 This is not the case in the dominantly spin singlet pairing state. 
 This is the main reason why the paramagnetic depairing effect in NCSC
depends on the symmetry of the leading order parameter. 
 We see that the \hc curves in Fig.~4 merge in the low magnetic field
region where the linear response theory is justified.

\begin{figure}[htbp]
  \begin{center}
\includegraphics[width=6.5cm]{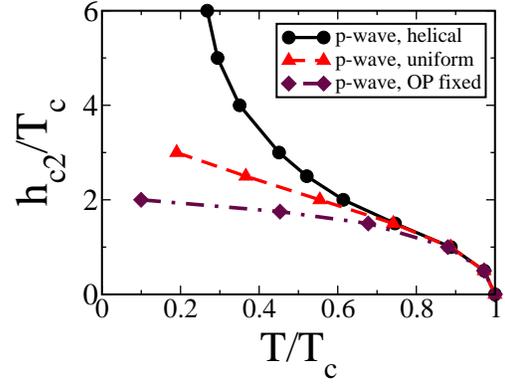}
\caption{(Color online)
 The circles (triangles) show the $H$-$T$ phase diagram 
in the helical (uniform) $p$-wave state. 
 We show the phase diagram in the SC state with 
$\vec{d}(\k)=(-\sin k_{\rm y}, \sin k_{\rm x},0)$, $ \Phi(\k) = 0$ 
and $\qfflo=0$ for a comparison. 
 The parameters are the same as in Fig.~4. 
}
  \end{center}
\end{figure}

 In order to shed light on  the mechanisms stabilizing the $p$-wave 
superconductivity at high magnetic fields, i.e., 
(i) the formation of helical SC state, and
(ii) the modification of SC order parameters, 
we compare \hc with the one for the uniform state with $\qfflo=0$ 
(triangles in Fig.~5) 
and the one for the SC state with 
$\vec{d}(\k)=(-\sin k_{\rm y}, \sin k_{\rm x},0)$, $\Phi(\k) = 0$ and 
$\qfflo=0$ (diamonds in Fig.~5). 
 Both (i) and (ii) are neglected in the latter (diamonds) while (i) is 
neglected in the former (triangles). The comparison between the 
triangles and diamonds shows the enhancement 
of \hc by optimizing the SC order parameter. 
Actually, the $d_{\rm x}$- ($d_{\rm y}$-)component of $d$-vector 
decreases in the magnetic field along the [100]- ([010]-)axis to 
avoid the paramagnetic depairing effect. 
 The \hc is furthermore enhanced below $T=0.6T_{\rm c}$ by forming the 
helical SC state (circles). 
 Thus, the $p$-wave superconductivity can be stabilized in the magnetic 
field which is much higher than the standard paramagnetic limit 
owing to the combination of mechanisms (i) and (ii).

\begin{figure}[htbp]
  \begin{center}
\includegraphics[width=6.5cm]{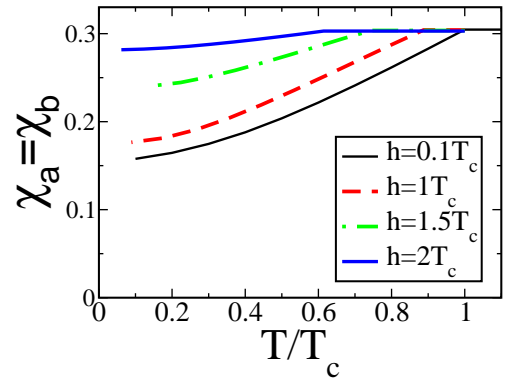}
\caption{(Color online)
The spin susceptibility along the [100]- and [010]-directions 
in the helical $p$-wave SC state without AFM order. 
 The magnetic field is chosen as $h=0.1 T_{\rm c}$, 
$h=T_{\rm c}$, $h=1.5 T_{\rm c}$ and $h=2 T_{\rm c}$ 
from the bottom to the top. 
The other parameters are the same as in Fig.~4. 
}
  \end{center}
\end{figure}

A large critical magnetic field \hc generally indicates that a SC state 
with a large spin susceptibility is stabilized at high magnetic fields. 
 The general spin susceptibility defined by
$\chi_{\rm a}=\chi_{\rm b}=M_{\rm x,y}/h$ is obtained from
the calculation of 
 the uniform magnetization, 
\begin{eqnarray}
\label{eq:magnetization-spin}
&& \hspace*{-5mm}
M_{\mu} = \sum_{k} {\rm Tr} \hat{S}^{\mu} \hat{G}(k), 
\end{eqnarray}
where $\hat{S}^{\mu}$ is the spin operator defined in eq.~(18). 
 The corresponding spin susceptibility for the $p$-wave state 
is shown in Fig.~6. 
 The spin susceptibility at $h=0.1T_{\rm c}$ drops to 
half of its normal state value at $T=0$. 
 This is consistent with the linear response theory in \S2.2. 
 For a magnetic field comparable to or higher than the standard 
paramagnetic limit, the order parameter of $p$-wave state is 
modified in order to avoid the paramagnetic depairing effect. 
 Therefore, 
the spin susceptibility at $h=2T_{\rm c}$ is almost constant through \Tcf, 
although the  critical temperature remains high, 
($T_{\rm c}(h=2T_{\rm c})=0.61T_{\rm c}(h=0)$). 
 These results should be contrasted to the $d_{\rm x^{2}-y^{2}}$-wave case 
discussed in Appendix B. The \Tc of $d_{\rm x^{2}-y^{2}}$-wave state 
is reduced more strongly ($T_{\rm c}(h=2T_{\rm c})=0.42T_{\rm c}(h=0)$), 
but the decrease of spin susceptibility below \Tc is larger than that 
in the $p$-wave state (Fig.~B.1).

\subsection{Role of anisotropic Fermi surface}

 A further important role in this context is played by the 
shape of the Fermi surface. 
 The band structure of the $\beta$-band is complicated but has one eye-catching
 property: 
 the cross sections of the Fermi surface in the range $k_{\rm z} > 2\pi/3$ 
are quadrilateral with the corners along the 
[110]- and [1-10]-directions.~\cite{rf:samokhinband,rf:yanaselett}   
 We show the schematic figures for the Fermi surface in Fig.~7 
where the anisotropy is stressed for simplicity. 
The anisotropy of the Fermi surface affects  \hc in two ways, by facilitating
(i) the formation of the helical SC phase and 
(ii) the modification of the SC order parameters, 
as we will discuss now.

\begin{figure}[htbp]
  \begin{center}
\includegraphics[width=6.5cm]{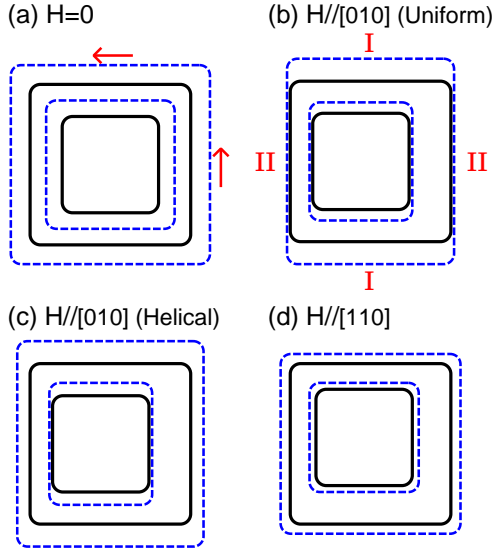}
\caption{(Color online)
The schematic figure for the Fermi surface and SC gap in the $p$-wave 
state. We show the cross section for a fixed $k_{\rm z}$. 
Two solid lines show the Fermi surfaces which are split 
by the ASOC. The dashed lines show the magnitude of SC gap on each 
Fermi surface. 
(a) The uniform BCS state at $\vec{H}=0$. The direction of $d$-vector 
is shown by the arrows. 
(b) The uniform BCS state for $\vec{H}//[010]$. 
The parts of Fermi surface ``I'' and ``II'' are shown. 
The SC gap on the part ``II'' is suppressed. 
(c) The helical SC state for $\vec{H}//[010]$. 
The SC gap on the part ``II'' of 
large Fermi surface is increased. 
(d) The helical SC state for $\vec{H}//[110]$. 
}
  \end{center}
\end{figure}

 First, (i) the helical SC phase is stable for the anisotropic
Fermi surface not only in the $p$-wave state but also 
in the $s$- and $d$-wave states. 
 This is simply because a set of quasi-particles with 
$\k =\pm \k +\qfflo/2$ can have low energy on a large part of 
the first Brillouin zone (nesting feature of the Fermi surface). 
 As shown in Fig.~7(b), the large (small) Fermi surface moves 
to the right (left) in the magnetic field along the [010]-axis. 
Under this condition, uniform Cooper pairing on the Fermi surface part parallel to [010]-axis 
(part ``II'' in Fig.~7(b)) is destabilized, while it is little affected on the part ''I''.  
 However, the depairing effect arising from the large Fermi surface 
is essentially avoided in the helical SC phase having 
$\qfflo \sim 2 h/v_{\rm F} \hat{x}$ (Fig.~7(c)) because of 
the nesting of Fermi surface along the [100]-direction. 
 This leads to the strong enhancement of \hcf. 
 This is not the case in the isotropic system 
where the Fermi surface is not nested.

 Second, the anisotropic Fermi surface enhances (ii) 
the mixing of order parameters 
and increases in this way \hc in the $p$-wave state. 
 Because of the structure of $g$-vector 
$ \vec{g}(\k) \propto (-v_{\rm y}(\k),v_{\rm x}(\k),0) $, 
the SC gap on the Fermi surface perpendicular to the [010]-axis 
(Fermi surface ``I'' in Fig.~7(b)) is mainly induced by the 
$d_{\rm x}$-component of spin triplet order parameter 
while the $d_{\rm y}$-component is the main source of the SC gap 
on the other part (Fermi surface ``II'' in Fig.~7(b)). 	
Since the $d_{\rm x}$- and $d_{\rm y}$-components induce the Cooper 
pairing on different parts of the Fermi surface, the coupling is weak 
between these two order parameters. 
Hence, the splitting of energy between
$\vec{d}(\k)=(-\sin k_{\rm y}, \sin k_{\rm x},0)$ (most stable state) and 
$\vec{d}(\k)=(\sin k_{\rm y}, \sin k_{\rm x},0)$ (second most stable state) 
due to the ASOC is small, and they can be easily mixed by the applied 
magnetic field.

\begin{figure}[htbp]
  \begin{center}
\includegraphics[width=7.5cm]{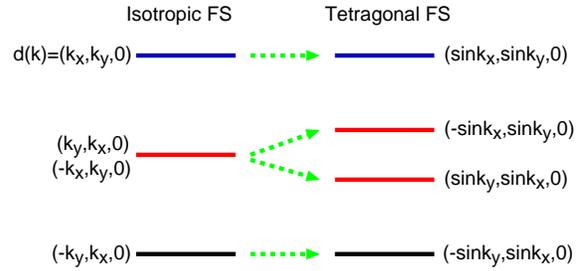}
\caption{(Color online)
 The schematic figure for the energy levels in the dominantly $p$-wave state. 
 The isotropic and tetragonal symmetries are assumed in the 
left and right figures, respectively.  
 The 2-fold degeneracy in the isotropic system between 
$\vec{d}(\k)=(k_{\rm y}, k_{\rm x},0)$ and 
$\vec{d}(\k)=(-k_{\rm x}, k_{\rm y},0)$ is lifted to 
$\vec{d}(\k)=(\sin k_{\rm y}, \sin k_{\rm x},0)$ and 
$\vec{d}(\k)=(-\sin k_{\rm x}, \sin k_{\rm y},0)$ in the tetragonal system. 
 In case of the $\beta$-band of \Ptf, 
$\vec{d}(\k)=(\sin k_{\rm y}, \sin k_{\rm x},0)$ has lower energy. 
}
  \end{center}
\end{figure}

In general, the tetragonal anisotropy of the Fermi surface reduces the 
splitting between the most stable and 
the second most stable pairing states. 
For an isotropic Fermi surface, the second most stable pairing state 
has 2-fold degeneracy; $\vec{d}(\k)=(k_{\rm y}, k_{\rm x},0)$ is 
degenerate with $\vec{d}(\k)=(- k_{\rm x}, k_{\rm y},0)$. 
 However, this degeneracy is lifted by the tetragonal anisotropy 
as shown in the schematic figure (Fig.~8).  
 This lift of degeneracy  decreases 
the difference of condensation energy between 
$\vec{d}(\k)=(-\sin k_{\rm y}, \sin k_{\rm x},0)$ and 
$\vec{d}(\k)=(\sin k_{\rm y}, \sin k_{\rm x},0)$ 
(or $\vec{d}(\k)=(-\sin k_{\rm x}, \sin k_{\rm y},0)$) in the system 
with tetragonal symmetry (see Fig.~8).

\begin{figure}[htbp]
  \begin{center}
\includegraphics[width=6.5cm]{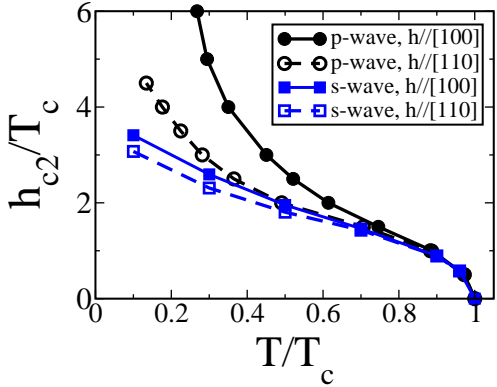}
\caption{(Color online)
 The critical magnetic field \hc along the [110]-axis 
in the helical $p$-wave (open circles) and $s$-wave (open squares) 
states.  
 Those along the [100]-axis are shown by the closed symbols for a 
comparison. 
}
  \end{center}
\end{figure}

Furthermore, a strong anisotropy of the Fermi surface induces a pronounced 
4-fold anisotropy in the paramagnetic properties. 
 Figure~9 shows that the \hc along the [110]-direction is much smaller than 
that along the [100]-direction in case of the $p$-wave state. 
 This is mainly because the state 
$\vec{d}(\k)=(-\sin k_{\rm y}, \sin k_{\rm x}, 0)$ is admixed 
by the magnetic field along the [110]-direction, with 
 $\vec{d}(\k)=(-\sin k_{\rm x}, \sin k_{\rm y}, 0)$, but not with 
 $\vec{d}(\k)=(\sin k_{\rm y}, \sin k_{\rm x}, 0)$. 
 The latter is less stable than the former in case of the $\beta$-band 
of \Ptf. 
On the other hand, the 4-fold anisotropy is weak for the $s$-wave state 
as shown by the squares in Fig.~9. 
 This indicates that the anisotropic Fermi surface enhances 
the \hc in the $p$-wave state mainly through the mixing of SC 
order parameters.

Finally, we comment on orbital depairing 
which we have neglected so far. 
 The orbital depairing effect is reduced by the mixing of 
order parameters in the $p$-wave state. 
 For example, the parameter $\beta$ in 
$\vec{d}(\k) = (-\sin k_{\rm y},\beta \sin k_{\rm x}, 
-{\rm i}\gamma \sin k_{\rm y})$ 
is decreased by the magnetic field along the [010]-direction,
and reduces the orbital depairing effect,  because the 
coherence length along the [100]-direction shrinks. 
 Thus, the \hc in the $p$-wave state is enhanced by modifying the 
order parameter through the suppression of the orbital depairing effect 
as well as the paramagnetic depairing effect.

\subsection{AFM state}

 In the discussion of the influence of AFM order on the 
$p$-wave SC state we focus on staggered moments along the [100]-axis with 
the magnetic field parallel to the [010]-axis, since the AFM moment 
favors to be perpendicular to the field. 
The situation of the magnetic field parallel to the moment is 
described in \S7. 

 The influence of the AFM order on the $p$-wave state significantly 
depends on the anisotropic spin-spin interaction, the $J$-term in eq.~(31). 
The critical field \hc depicted in Fig.~10  with $h_{\rm Q}=0.125$ 
in the AFM ordered phase shows a clear trend. 
 While the AFM leads to a reduction of \hc in 
the absence of the anisotropic spin interaction ($J=0$), 
a strong enhancement is obtained for
$ J=0.3 $.

\begin{figure}[htbp]
  \begin{center}
\includegraphics[width=6.5cm]{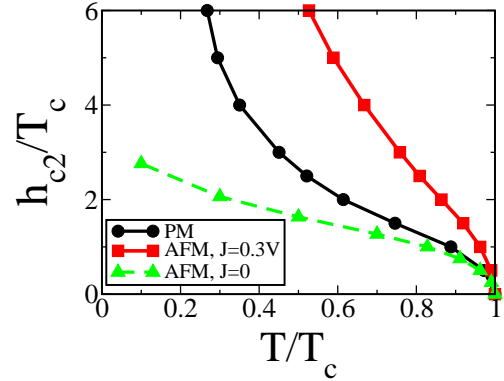}
\caption{(Color online)
The $H$-$T$ phase diagram in the $p$-wave state for the magnetic field 
along the [010]-axis in the presence of AFM moment along the [100]-axis. 
 The squares and triangles show the \hc for $J=0.3V$ and $J=0$, 
respectively. 
 We fix $h_{\rm Q}=0.125$ and choose the other parameters as in Fig.~4. 
 The \hc in the PM state is shown for a comparison (circles).  
}
  \end{center}
\end{figure}

 We understand these results by analyzing the parameter $\beta$ of 
$\vec{d}(\k) = (-\sin k_{\rm y},\beta \sin k_{\rm x}, 0)$ at zero 
magnetic field.

 For $\beta < 1$, the superconductivity is dominant on 
Fermi surface region ``I'' in Fig.~7(b), while the magnetic field along the 
[010]-axis suppresses Cooper pairing on the Fermi surface ``II''. 
For this reason, this SC state is robust against the magnetic field 
along the [010]-axis. The magnetic field reduces $\beta$ even more 
enhancing the anisotropy of the SC gap. 
 The enhancement of \hc due to the AFM order is much more significant 
than expected in the simple estimation eq.~(42). 
 In fact, the suppression of paramagnetic depairing effect in case of 
$\beta <1$ can be viewed as a result of the strong non-linear response 
to the magnetic field. 
 The small energy scale $\beta T_{\rm c}$ appears in this case and 
induces the strong non-linearity. 
 This is the reason why the influence of AFM order is much more important 
in the $p$-wave state than in the $s$-wave state.  If we assume $J=0$, the parameter $\beta$ is more than unity, 
which is incompatible with our RPA analysis for the Hubbard 
model.~\cite{rf:yanaselett} 
 On the other hand, we obtain $\beta \sim 0.3$ for $J=0.3 V$ and 
$h_{\rm Q}=0.125$, giving the 
result consistent with the RPA theory.

\begin{figure}[htbp]
  \begin{center}
\includegraphics[width=6.5cm]{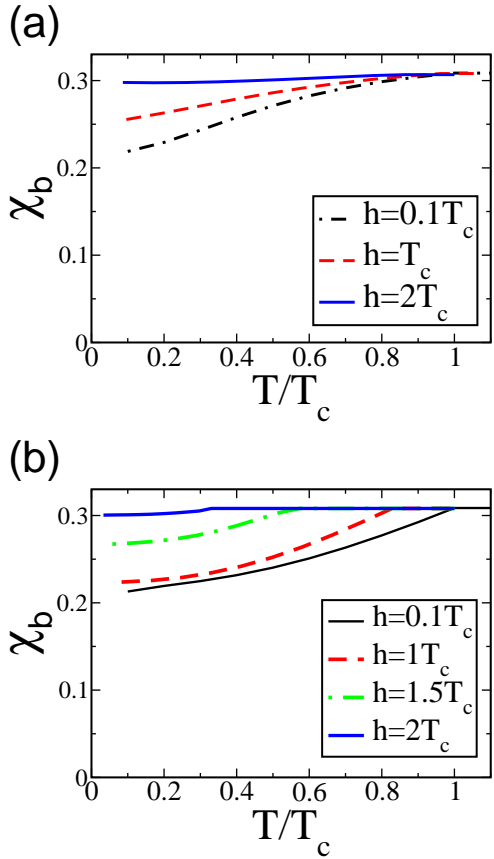}
\caption{(Color online)
The spin susceptibility along the [010]-direction in the helical 
$p$-wave state with AFM order. 
 We assume $J=0.3 V$ leading to $\beta < 1$ in (a) and 
$J=0$ leading to $\beta > 1$ in (b), respectively.  
The magnetic field is shown in the figures. 
The other parameters are the same as in Fig.~10. 
}
  \end{center}
\end{figure}

 The strong non-linear response to the magnetic field clearly appears  
in the magnetic field dependence of spin susceptibility. 
 We show the spin susceptibility $ \chi_{\rm b} $ for $J=0.3V$ ($\beta < 1$) 
and $J=0$ ($\beta > 1$) in Figs.~11 (a) and (b), respectively. 
For low magnetic fields ($h=0.1T_{\rm c}$) $ \chi_{\rm b} $ is enhanced by
AFM order in both cases consistent with the linear response theory (\S2.4). 
We find that $ \chi_{\rm b}$ is furthermore enhanced for the moderate 
magnetic field $h=T_{\rm c}$ with $\beta < 1$ (Fig.~11(a)) 
although the critical temperature $T_{\rm c}(h)$ is little decreased. 
 According to these theoretical results, the NMR Knight shift 
measurement~\cite{rf:yogiK} by Yogi {\it et al.} and 
the $\mu$SR measurement~\cite{rf:higemoto} by Higemoto {\it et al.} 
were carried out in the non-linear response regime.

 In contrast to $\beta <1 $, the moderate magnetic field 
$h=T_{\rm c}$ little affects the spin susceptibility (Fig.~11(b)). 
 The non-linearity of spin susceptibility appears only in the high field 
region close to the critical magnetic field. 
 This is a characteristic property of the SC state with strong 
paramagnetic depairing effect such as the spin singlet pairing state 
in centrosymmetric system. 
 Qualitatively the same magnetic field dependence is obtained  
in the dominantly $d_{\rm x^{2}-y^{2}}$-wave state (see Fig.~B.1 in 
Appendix B).

\section{Pairing symmetry in CePt$_3$Si}

Measurements of  \Hc and the Knight shift are consistent with $p$-wave
superconductivity in CePt$_3$Si at ambient pressure. 
 The temperature dependence of \Hcf~\cite{rf:bauerDC,rf:bauerreview,
rf:onukireview,rf:yasuda} implies the absence of paramagnetic depairing. 
NMR and $\mu$SR Knight shift data show no decrease below 
\Tcf,~\cite{rf:yogiK,rf:higemoto} although \Tc remains rather high at 
applied magnetic fields. These findings could be understood based on
the $p$-wave state with AFM order for which the theoretical 
results have been shown in Figs.~10 and 11(a).

 We here note that the other possible mechanisms for the high critical 
field $H_{\rm c2}$ are unlikely relevant in \Ptf. 
 For example, a small $g$-factor has been suggested for CeCoIn$_5$ 
($g \sim 0.63$).~\cite{rf:miclea} 
 It is expected that the $g$-factor of CeCoIn$_5$ is significantly 
renormalized by the strong AFM correlation in the 
{\it ab}-plane.~\cite{rf:yanaseFFLO} 
 However, this is not the case in \Pt where the spin correlation in the 
{\it ab}-plane are dominantly ferromagnetic.~\cite{rf:yanaselett,rf:metoki} 
 The strong coupling effect which has been ignored in this paper is 
another possible cause of high $H_{\rm c2}$. 
 But, the jump of the specific heat at $T=T_{\rm c}$ 
does not indicate strong coupling effects 
in \Ptf,~\cite{rf:bauerDC,rf:bauerreview,rf:onukireview,rf:takeuchiC} 
in contrast to \Irf.~\cite{rf:tateiwastrongcoupling}

 The $p$-wave state is consistent with the coherence peak in NMR 
$1/T_{1}T$~\cite{rf:yogiT1,rf:hayashiT1,rf:fujimoto} 
and the line node behaviors in various 
quantities.~\cite{rf:bonalde,rf:izawa,rf:takeuchiC,rf:yanaselett,
rf:fujimotoGap,rf:hayashiSD} Moreover the microscopic theory 
within an RPA theory suggests an in-plane $p$-wave state 
induced by the $\beta$-band of \Ptf.~\cite{rf:yanaselett}

For $\vec{H} \perp \vec{h}_{\rm Q}$ 
the experimental magnetic properties of \Pt at ambient pressure 
are consistent with the $p$-wave state with $\beta <1$. This indicates the strong anisotropy of 
the effective spin interaction, which is described by the $J$-term in eq.~(31)
and is compatible with the RPA analysis.~\cite{rf:yanaselett} It does however 
not agree with the naive second order perturbation theory 
which leads to the $p$-wave state with $\beta > 1$.~\cite{rf:yanasefull} 
 This is because the role of spin fluctuation is underestimated 
within the perturbation theory.~\cite{rf:yanaseReview} 
Based on this fact we may state that there is some evidence for 
spin-fluctuation-mediated superconductivity in CePt$_3$Si.

 When the magnetic field is parallel to the AFM moment 
$\vec{H} \parallel \vec{h}_{\rm Q}$, the paramagnetic depairing effect 
is enhanced (suppressed) in the $p$-wave state with $\beta <1$ ($\beta >1$). 
 We have confirmed that the \hc for $\vec{H} \parallel \vec{h}_{\rm Q}$ 
with $\beta >1$ is qualitatively the same as the \hc for 
$\vec{H} \perp \vec{h}_{\rm Q}$ with $\beta <1$ (squares in Fig.~10). 
 If the sample had a domain structure with respect to the direction of 
AFM moment, the SC state with the maximal \Tc would mark the
SC transition. Under such circumstances $p$-wave states with both 
$\beta > 1$ and $\beta < 1$ could ''avoid'' 
the paramagnetic depairing effect and would be consistent with the 
experimental results in \Ptf.~\cite{rf:yogiK,rf:higemoto,rf:bauerDC,
rf:bauerreview,rf:onukireview,rf:yasuda}

We here comment on the inter-plane $d$-wave state which we found as another 
possible pairing state on the basis of the RPA theory.~\cite{rf:yanaselett}
 Although the 2-fold degeneracy exists in this state 
($d_{\rm xz}$- and $d_{\rm yz}$-wave), 
the order parameter has no internal degree of freedom with respect to 
the spin. Therefore, the paramagnetic depairing effect cannot be avoided 
by modifying the order parameter in contrast to the $p$-wave state. 
Hence, the magnetic properties are qualitatively the same as those in 
the $s$-wave state 
which seem to be incompatible with the experimental results in CePt$_3$Si. 
 The inter-plane $d$-wave state is incompatible with the coherence peak 
in the NMR $1/T_{1}T$ too.~\cite{rf:yogiT1}

\section{Proposals for test experiments}

Here we discuss several experiments which could help to establish
the pairing symmetry for CePt$_3$Si as well as CeRhSi$_3$ and CeIrSi$_3$. 

The influence of antiferromagnetism on the magnetic properties can be tested by using the
fact that AFM order can be suppressed by pressure in these materials.~\cite{rf:tateiwa,rf:takeuchiP,rf:sugitani,rf:onukireview,
rf:kimura,rf:kimurareview} It follows from our results in \S4, \S5.1 and 
Appendix B, 
that in the purely SC phase the paramagnetic depairing should limit
the upper critical field for $\vec{H} \parallel ab$ and the spin susceptibility
should decrease below \Tc in the low-magnetic field regime. 
 Actually recent measurements of $H_{\rm c2}$ along {\it ab}-plane in 
CeRhSi$_3$~\cite{rf:kimurareview,rf:kimuraprivate} 
and CeIrSi$_3$~\cite{rf:okuda,rf:onukireview,rf:onukiprivate} 
imply a clear paramagnetic depairing effect in the purely SC region, 
consistent with the theoretical view. 
 Notably paramagnetic depairing seems less 
effective in the AFM state of \Irf.~\cite{rf:onukiprivate} This is compatible
with $p$-wave pairing. No studies of this kind have been performed so
far for \Ptf. 

 A further aspect is the  2-fold in-plane anisotropy 
in the AFM state. Since the [100]- and [010]-axes are not equivalent 
in the AFM state, 
a 2-fold anisotropy appears in the {\it ab}-plane. 
 Although the AFM moment perpendicular to the uniform magnetic field 
is generally favored, the situation $\vec{H} \parallel \vec{h}_{\rm Q}$ 
can nevertheless be realized for magnetic fields low enough to leave
 the orientation of the AFM moment unchanged. 

 We summarize the 2-fold anisotropy expected for each pairing state 
in Fig.~12 taking also the orbital depairing effect  into account. 
We assumed here that the $H_{\rm c2}$ determined by the orbital depairing 
is much higher than the standard paramagnetic limit field
in CePt$_3$Si,~\cite{rf:bauerDC,rf:bauerreview,rf:onukireview,rf:yasuda} 
CeRhSi$_3$~\cite{rf:kimura,rf:kimurareview} 
and CeIrSi$_3$.~\cite{rf:okuda,rf:onukireview} 
The upper critical field due to orbital depairing is naturally enhanced by the heavy mass of 
quasi-particles in these heavy Fermion compounds. Under such 
conditions paramagnetic depairing can play a role in the high-field regime.

Fig.~12(a) shows the $H$-$T$ phase diagram in the $p$-wave state. 
For $\vec{H} \parallel \vec{h}_{\rm Q}$ the paramagnetic depairing effect 
is enhanced (suppressed) with $\beta < 1$ ($\beta > 1$). Note that the opposite
occurs for $\vec{H} \perp \vec{h}_{\rm Q}$ (see Fig.~10). 
 Therefore, a significant 
2-fold anisotropy of $H_{\rm c2}$ could appear
at high magnetic fields for either $\beta < 1$ or $\beta > 1$, provided 
the AFM moment remains pinned. 
 Qualitatively the same anisotropy would occur at low magnetic fields, 
because the orbital depairing effect is anisotropic owing to the 
in-plane anisotropy of coherence length, namely the difference of 
$\xi_{\rm a}$ and $\xi_{\rm b}$. 
On the basis of the 
RPA theory for \Ptf~\cite{rf:yanasefull} we have estimated the anisotropy as 
$\partial H_{\rm c2}/\partial T|_{T=T_{\rm c}} (\vec{H} \parallel [100]): 
\partial H_{\rm c2}/\partial T|_{T=T_{\rm c}} (\vec{H} \parallel [010])= 
\xi_{\rm a}: \xi_{\rm b} = 0.672 :1$ at $h_{\rm Q}=0.125$. 
 Thus, the $H$-$T$ phase diagram is highly anisotropic in both 
high and low magnetic field region as shown in Fig.~12(a).

\begin{figure}[htbp]
  \begin{center}
\includegraphics[width=7cm]{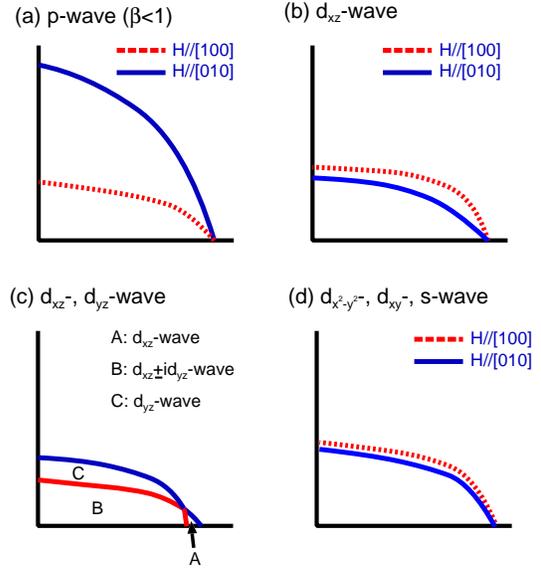}
\caption{(Color online)
Schematic figure for the 2-fold anisotropy in the $H$-$T$ phase diagram. 
We assume the AFM order along the [100]-axis. 
(a) The $p$-wave state with $\beta < 1$. 
The opposite anisotropy is expected for $\beta > 1$. 
(b) The inter-plane $d$-wave ($d_{\rm xz}$-wave) state. 
(c) Possible multiple phase transitions in the inter-plane $d$-wave state 
for $\vec{H} \parallel [010]$. 
(d) The intra-plane $d$-wave ($d_{\rm x^{2}-y^{2}}$-wave or $d_{\rm xy}$-wave) 
and $s$-wave states. 
}
  \end{center}
\end{figure}

 The strong 2-fold anisotropy in the {\it ab}-plane appears also 
in the inter-plane $d$-wave state due to the anisotropy of 
the coherence length. The 2-fold degeneracy between the $d_{\rm xz}$- and $d_{\rm yz}$-wave states 
is lifted by the AFM order. The staggered moment along the [100]-axis favors the 
$d_{\rm xz}$-wave state and yields a coherence length which is longer along the [100]-axis than 
along the [010]-axis. For this reason \Hc close to $T=T_{\rm c}$ is smaller for the 
magnetic field along the [010]-axis. This anisotropy is suppressed at high magnetic fields 
because the  paramagnetic depairing effect is nearly isotropic 
as in the $s$-wave state (Fig.~3). 
These considerations lead to the schematic phase diagram in Fig.~12(b).

 It should be noted that the in-plane anisotropy of \Hc in the inter-plane 
$d$-wave state does not vanish if  the quantum critical point 
of the AFM order is approached. 
 This is in contrast to the $p$-wave state where the in-plane anisotropy 
is suppressed by decreasing the AFM moment. 
 In the vicinity of AFM quantum critical point, 
multiple phase transitions can occur for the inter-plane $d$-wave state 
as discussed in Ref.~27. 
 These multiple phases in the $H$-$T$ plane are shown in Fig.~12(c) for 
the magnetic field along the [010]-axis. 
Pure $d_{\rm xz}$- and $d_{\rm yz}$-wave states appear
in the high-temperature region and in the high-magnetic field region, 
respectively. 
The chiral $d_{\rm xz} \pm {\rm i}d_{\rm yz}$-wave state 
is stabilized at low temperatures and  fields. 
 If the multiple phase transitions were observed in the $H$-$T$ plane or 
in the $P$-$T$ plane, it would be a strong evidence for the inter-plane 
$d$-wave state. 
 Although some indications for a second SC transition have been reported 
in CePt$_3$Si,~\cite{rf:scheidt,rf:nakatsuji,rf:bonalde,rf:aoki} 
it remains unclear 
whether it represents an intrinsic property or is caused by
the sample inhomogeneity.

 In contrast to the $p$-wave and inter-plane $d$-wave states, 
the 2-fold anisotropy of \Hc is very weak in the intra-plane 
$d$-wave and $s$-wave states because the paramagnetic depairing effect 
as well as the orbital depairing effect are nearly isotropic. 
 Therefore, we obtain a simple phase diagram in Fig.~12(d). 

 Since the 2-fold anisotropy of \Hc is quite different between the dominantly 
$p$-wave, inter-plane $d$-wave and intra-plane spin singlet pairing states, 
the future experiment in the AFM state could identify the pairing symmetry 
in \Ptf, \Rh and \Irf. 
 It should be noticed that this experiment can be performed 
in \Pt without applying the pressure.

\section{Helical Superconductivity}

 In this section we discuss the nature of the helical SC state 
which is a novel SC phase specific to NCSC. 
 The SC phase with a finite total momentum of Cooper pairs 
$\qfflo$ is stabilized in the presence of Rashba-type 
spin-orbit coupling under a magnetic field in the
{\it ab}-plane.~\cite{rf:samokhinHelical,rf:kaurHV,
rf:agterberg,rf:oka,rf:mineevHelical,rf:feigelman} 
This state bears some similarity with the FFLO state~\cite{rf:FFLO} 
in centrosymmetric superconductors, but has also important 
differences.
 First, the helical SC phase is stabilized immediately above $H_{\rm c1}$ which 
is much lower than $H_{\rm c2}$ in extremely type II superconductors. 
 This is in contrast to the FFLO state which appears in a
narrow region near $H_{\rm c2}$ only. 
 Second, the phase of SC order parameter is modulated as 
$\Delta(\vec{r})=\Delta e^{{\rm i} \qfflo \vec{r}}$ in the helical SC state 
(which is the same form as in the Fulde-Ferrel (FF) state) while the 
Larkin-Ovchinnikov (LO) state with the 
spatial modulation of the amplitude,
$\Delta(\vec{r})=\Delta \cos\qfflo\vec{r}= 
\Delta (e^{{\rm i} \qfflo \vec{r}} + e^{-{\rm i} \qfflo \vec{r}})/2$, 
is more stable than the FF state.~\cite{rf:matsudareview} 
 Because the two momenta $\qfflo$ and $-\qfflo$ are equivalent in the 
centrosymmetric system, the order parameter has a double $q$ structure 
in the LO state. 
 On the other hand, $\qfflo$ is not equivalent to $-\qfflo$ in 
the non-centrosymmetric system under a magnetic field. 
For this reason the helical SC phase is realized in the NCSC
at least just below the critical temperature. At higher fields and low temperature
also a  ``stripe SC state'' can be realized,~\cite{rf:agterberg,rf:feigelman} 
which is similar to the LO state.

Experimental evidence for the FFLO state has been obtained for  
CeCoIn$_5$~\cite{rf:kumagai}, more than forty years after the 
theoretical proposal.~\cite{rf:FFLO,rf:matsudareview} 
 This is partly because the FFLO state is suppressed 
by a weak disorder.~\cite{rf:adachi} 
 The stripe SC state, which is resembles the FFLO state,
can be suppressed by a weak disorder too. 
 In contrast to these states the helical SC state is realized 
even in the disordered material, if the superconductivity is present. 
 Although there is no experimental verification of the helical SC state in 
NCSC so far, the existence of the helical SC phase 
is a mandatory features from a theoretical point of view.

Now we turn to the effect of finite $\qfflo$ on the paramagnetic 
properties. 
 Although the influence of the helical superconductivity has been taken into 
account in \S4 and \S5, the following discussion will be important for a 
deeper understanding.

 One of the characteristic properties in the helical SC state 
is the presence of a finite spin magnetization. 
In the low magnetic field region this magnetization is expressed  as 
$\vec{M}=\vec{M}_{\rm 0}+\hat{\chi}' \vec{H}$ with finite $\vec{M}_{\rm 0}$. 
 For simplicity, we here consider the PM state 
and assume the SC order parameter without gap nodes. 
 Then, the magnetization is obtained as, 
\begin{eqnarray}
\label{eq:pauli-Yosida-Helical-para-full}
&& \hspace*{-10mm}
\vec{M}_{0} = \frac{1}{4} \sum_{\k} \vec{\tilde{g}}(\k) 
(\vec{B}_{1}(\k) \cdot \qfflo - \vec{B}_{2}(\k) \cdot \qfflo) 
\\ 
&& \hspace*{-10mm}
\sim 
\frac{1}{2} 
[\int {\rm d}\k_{\rm F} \vec{\tilde{g}}(\k_{\rm F})
(\vec{v}_{1}(\k_{\rm F}) \cdot \qfflo) /v_{1}(\k_{\rm F}) 
- (1 \leftrightarrow 2)]
\\ && \hspace*{-10mm}
= D (\hat{z} \times \qfflo), 
\end{eqnarray}
with $D \propto \alpha$. We define
$\vec{B}_{\gamma}(\k)={\rm d}(\ee_{\gamma}(\k)/E_{\gamma}(\k))/{\rm d}\k$ 
where $\gamma$ is a band index.
 As shown in eq.~(46), the magnetization is oriented
along the direction perpendicular to $\qfflo$. 

 The helical superconductivity also affects the differential 
spin susceptibility $\chi'_{\mu\mu}={\rm d} M_{\mu}/{\rm d} H_{\mu}$ 
when the SC gap has a node. 
 According to eqs.~(11-13), the quasiparticles suffer a Doppler 
shift~\cite{rf:volovik} in the helical SC state and 
the single particle excitation energy is expressed as 
$\sqrt{\ee_{\gamma}(\k)^{2}+|\Delta_{\gamma}(\k)|^{2}}  
\pm \vec{v}_{\gamma}(\k) \cdot \qfflo/2$. 
 Following eq.~(15), the Pauli part of differential spin susceptibility 
is obtained as, 
\begin{eqnarray}
\label{eq:pauli-Yosida-Helical}
&& \hspace*{-8mm}
\chi_{\mu\mu}^{'\rm P}=
\sum_{\gamma} \int {\rm d}\k_{\rm F} 
A^{\mu\mu}_{\gamma\gamma}(\k_{\rm F}) 
\nonumber \\ &&
\times 
Y_{\rm H}(\vec{v}_{\gamma}(\k_{\rm F}),|\Delta_{\gamma}(\k_{\rm F})|,T)/
v_{\gamma}(\k_{\rm F}), 
\end{eqnarray}
for $\mu=x,y$ where $Y_{\rm H}(\vec{v},\Delta,T)$ is the generalized 
Yosida function, 
\begin{eqnarray}
\label{eq:Yosida-Helical}
&& \hspace*{-10mm}
Y_{\rm H}(\vec{v},\Delta,T)=- \frac{1}{2} \int {\rm d}\e 
[f'(\sqrt{\e^{2}+\Delta^{2}}+\vec{v}\cdot\qfflo/2) 
\nonumber \\ && \hspace*{23mm}
+ f'(\sqrt{\e^{2}+\Delta^{2}}-\vec{v}\cdot\qfflo/2)].  
\end{eqnarray}
 Since $Y_{\rm H}(\vec{v},\Delta,0) = 
\frac{1}{2} \sqrt{1-4 \Delta^{2}/(\vec{v}\cdot\qfflo)^{2}}$ 
for $|\Delta| < |\vec{v}\cdot\qfflo|/2$, 
the Doppler shift boosts the differential spin susceptibility in the SC state 
with a gap node like in CePt$_3$Si.~\cite{rf:izawa,rf:bonalde,rf:takeuchiC}

 In fact, the uniform BCS state is favored at $H=0$ 
and the helical SC state is induced by an infinitesimal magnetic 
field owing to the linear coupling between the magnetization and the 
helicity $\qfflo$ (eq.~(46)) with $\qfflo \perp \vec{H}$. Since the amplitude of $\qfflo$ 
is linear in the magnetic field $|\vec{H}|$, the formation of helical 
SC state leads to a correction to the linear response theory in \S2 of the uniform state. 
However, the correction is negligible when $|\alpha| \ll \e_{\rm F}$ 
because the amplitude of $\qfflo$ is small,  
$|\qfflo| \sim (\alpha/\e_{\rm F}) h/v_{\rm F}$ in linear order of 
small parameter $\alpha/\e_{\rm F}$.

The helicity can play a quantitatively more important role 
in the non-linear response regime, 
because the amplitude of $\qfflo$ increases from 
$|\qfflo| \sim (\alpha/\e_{\rm F}) h/v_{\rm F}$ in the low field region 
to $|\qfflo| \sim h/v_{\rm F}$ in the high field region 
with a rapid crossover around $h \sim T_{\rm c}$. 
 For example, Fig.~13 shows the magnetic field dependence of the helicity 
$|\qfflo|$ in the $p$-wave state, with a sharp increase of the helicity above $h = T_{\rm c}$.
As a result the critical field \hc is significantly enhanced 
at high fields as shown in Figs.~2 and 5.

\begin{figure}[htbp]
  \begin{center}
\includegraphics[width=6.5cm]{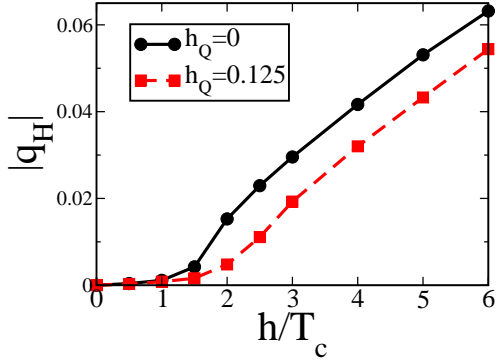}
\caption{(Color online)
 The amplitude of the helicity $|\qfflo|$ just below $T_{\rm c}$. 
 The circles and squares show the results in the PM and AFM states, 
respectively. 
 We assume the magnetic field along the [010]-direction which leads to 
$\qfflo$ along the [100]-direction. $J=0.3V$ is assumed in the AFM state. 
The other parameters in the PM and AFM states are the same as in 
Figs.~4 and 10, respectively. 
}
  \end{center}
\end{figure}

 The nature of the crossover from $|\qfflo| \sim (\alpha/\e_{\rm F}) h/v_{\rm F}$ 
to $|\qfflo| \sim h/v_{\rm F}$ becomes obvious viewing the
momentum dependence of eigenvalues $\lambda(\vec{q})$ in eqs.~(40) and (41). 
 Figures~14(a) and (b) show the numerical results in the PM and AFM states, 
respectively. 
 In Fig.~14(a), $\lambda(\vec{q})$ possesses a crossover from a single
to a double peak structure, yielding a rapid increase of the helicity. 
 This result implies that the nature of the helical SC phase is different 
below and above the crossover magnetic field. 
 Actually, the ``stripe SC state'' can be stabilized 
above the crossover field.~\cite{rf:agterberg} 
 As shown in Fig.~14(b), the crossover from the single to the double peak structure 
 is suppressed by the AFM order. 
 The eigenvalue $\lambda(\vec{q})$ has a single peak even in the 
magnetic field much higher than the standard paramagnetic limit. 
 This is simply because the AFM order suppresses the paramagnetic 
depairing effect in the $p$-wave state. 
 
We would like to point here
that \Pt is a good candidate for an experimental 
observation of the helical SC phase. Actually, the large critical field 
$H_{\rm c2}$ leads to the helical SC phase with large $\qfflo$ 
($|\qfflo| \sim h/v_{\rm F}$) 
in a large part of the $H$-$T$ phase diagram. 
 It seems to be difficult to detect the helical SC phase with small 
helicity $|\qfflo| \sim (\alpha/\e_{\rm F}) h/v_{\rm F}$ because the  
wave length is much longer than the coherence length. 
 Thus the high field phase with $|\qfflo| \sim h/v_{\rm F}$ is more promising 
for the experimental observation. 
 The high field phase is stable in the $p$-wave state 
above $h = T_{\rm c}$ as shown in Figs.~4, 10 and 13. 
 However, this phase shrinks in the SC state with dominantly spin singlet 
pairing and/or the strong orbital depairing effect which leads to 
small $H_{\rm c2}$.

\begin{figure}[htbp]
\begin{center}
\includegraphics[width=6.5cm]{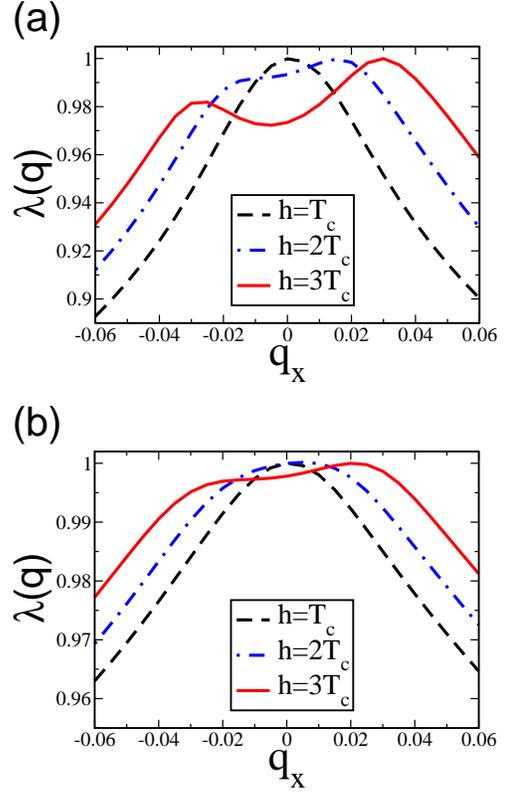}
\caption{(Color online)
The momentum dependence of the eigenvalue $\lambda(\vec{q})$ in the 
linearized mean field equation (eqs.~(40) and (41)) with 
$\vec{q}=(q_{\rm x}, 0, 0)$. 
(a) the PM state and (b) the AFM state.
The parameters are the same as in Fig.~13. 
}
  \end{center}
\end{figure}

\section{Summary and Discussions}

We have investigated the paramagnetic properties in NCSC. 
The SC states with leading $p$-wave, $d$-wave or $s$-wave 
order parameter have been examined in view of the
heavy Fermion superconductors, CePt$_3$Si, CeRhSi$_3$ and CeIrSi$_3$.

 First, the linear response to the magnetic field has been investigated 
with the particular interest on the role of AFM order. 
 The spin susceptibility is universal in the sense that it is 
independent of the pairing symmetry at $T=0$, if the ASOC is 
much larger than the SC gap, and results from the band splitting 
due to the ASOC. 
 The spin susceptibility below \Tc is increased in the AFM state 
due to the folding of unit cell, if the magnetic field is 
applied perpendicular to the AFM moment. 
 The result is opposite for the magnetic field parallel to the 
AFM moment.

 Second, we have shown that the non-linear response to the magnetic field 
depends on the symmetry of the leading SC order parameter.
 In particular, the spin susceptibility and $H_{\rm c2}$ for the $p$-wave 
state are significantly enhanced by non-linear effects through  
(i) the formation of a helical SC state and 
(ii) the mixing of SC order parameters. 
The anisotropy of Fermi surface can increase these non-linear effects,
and strengthen the influence of AFM order. 
Taking these aspects into account, the experimental 
results~\cite{rf:yogiK,rf:higemoto,rf:yasuda,rf:bauerDC,
rf:bauerreview,rf:onukireview} for \Pt 
at ambient pressure (within the AFM state) are consistent with 
the $p$-wave state admixed with a secondary $s$-wave component. 
This is the pairing state which has been proposed recently by 
Frigeri {\it et al.}~\cite{rf:frigeri} and identified by the 
microscopic RPA theory.~\cite{rf:yanaselett,rf:yanasefull} 
 It has been shown that this $p$-wave state is consistent with 
the line node behavior~\cite{rf:yanaselett,rf:yanasefull,
rf:bonalde,rf:izawa,rf:takeuchiC} and also with the coherence peak in 
NMR $1/T_{1}T$.~\cite{rf:hayashiT1,rf:fujimoto,rf:yogiT1}
 Although the RPA theory has identified the inter-plane $d$-wave 
($d_{\rm xz}$- and $d_{\rm yz}$-wave) state as further candidate, 
this state seems to be incompatible with the Knight shift, 
$H_{\rm c2}$ and NMR $1/T_{1}T$ measurements at ambient pressure. 
 According to these comparisons between the theory and experiment, 
CePt$_3$Si is rather likely the first identified 
spin triplet superconductor in Ce-based 
heavy fermion systems.

 We have proposed several experiments which can provide further
evidences for the pairing state in CePt$_3$Si as well as in \Rh and \Irf. 
The first proposal is the pressure dependence in various quantities. 
If the AFM order is a major cause of the unusual properties in CePt$_3$Si, 
a pronounced pressure dependence is expected in NMR, specific heat, 
thermal transport, superfluid density and so on.  
 If CePt$_3$Si has a leading $p$-wave order parameter, the following 
behaviors are expected above the critical pressure $P \sim 0.6$GPa.  
(a) The Knight shift decreases below \Tc for the magnetic field along the 
{\it ab}-plane and below the standard paramagnetic limit. 
(b) The paramagnetic depairing effect is enhanced for 
$\vec{H} \parallel${\it ab} but not for $\vec{H} \parallel${\it c}. 
(c) The low-energy excitations due to the accidental line nodes are 
decreased.~\cite{rf:yanaselett} 
(d) The coherence peak in NMR $1/T_{\rm 1}T$ is enhanced by the 
isotropic SC gap.~\cite{rf:yanasefull} 
 These pressure dependences are not expected in the intra-plane $d$-wave 
($d_{\rm x^{2}-y^{2}}$- and $d_{\rm xy}$-wave) and $s$-wave states. 
 The pressure dependence (c) is expected also in the inter-plane $d$-wave 
state and then the additional phase transition occurs 
in the $P$-$T$ and $H$-$T$ plane.~\cite{rf:yanaselett}

 Another proposal for a future experiment is the 2-fold anisotropy 
arising from the AFM order. 
 The strong 2-fold anisotropy is expected in the dominantly $p$-wave state 
while the anisotropy is negligible in the $s$-wave and intra-plane $d$-wave 
states. 
 In the inter-plane $d$-wave state the strong 2-fold anisotropy 
is expected near $T=T_{\rm c}$ but the anisotropy is suppressed 
at high magnetic fields. 
 The experimental observation of the 2-fold anisotropy in the AFM state 
could provide an important evidence for the pairing symmetry 
in \Ptf, \Rh and \Irf. 
 Thus, the response to the AFM order can be a signature of the pairing 
symmetry in non-centrosymmetric superconductors.

\section*{Acknowledgments}

 The authors are grateful to D. F. Agterberg, J. Akimitsu, 
J. Flouquet, P.A. Frigeri, S. Fujimoto, J. Goryo, 
N. Hayashi, K. Izawa, N. Kimura, Y. Kitaoka, Y. Matsuda, 
V. P. Mineev, H. Mukuda, 
E. Ohmichi, Y. \Onuki, T. Shibauchi, R. Settai, T. Takeuchi, H. Tanaka, 
T. Tateiwa, H. Tou and M. Yogi 
for fruitful discussions. 
 This study has been financially supported by the Nishina Memorial 
Foundation, Grants-in-Aid for Young Scientists (B) 
from the Ministry of Education, Culture, Sports, 
Science and Technology (MEXT), 
the Swiss Nationalfonds and the NCCR MaNEP. 
 Numerical computation was carried out 
at the Yukawa Institute Computer Facility.

\appendix

\section{Linear Response Theory}

 The dynamical spin susceptibility in the linear response regime 
is obtained by the Kubo formula as, 
\begin{equation}
\label{eq:spin-susceptibility}
\begin{array}{l}
\chi_{\mu\nu}(q)= \\  \\
 \quad  \displaystyle -\sum_{\gamma,\delta} \sum_{k}{'} 
[S^{\mu}_{\delta\gamma}(\k+\q,\k) S^{\nu}_{\gamma\delta}(\k,\k+\q)
G_{\delta}(k+q)G_{\gamma}(k)  \\
\qquad -  S^{\mu}_{\gamma\delta}(\kp+\q,\kp) S^{\nu}_{\gamma\delta}(-\km-\q,-\km) 
  F_{\delta}(k+q)F^{\dag}_{\gamma}(k)].
\end{array}
\end{equation}
where $q=(\q,{\rm i}\Omega_{n})$, $k=(\k,{\rm i}\omega_{n})$
and $\q$ is the momentum along the {\it ab}-plane. 
 The spin operator $S^{\mu}_{\gamma\delta}(\k+\q,\k)$ 
in the band basis has been given in eq.~(17). 

 Taking the limit $\Omega_{n} \rightarrow 0$ and 
$\q \rightarrow 0$, 
we obtain the uniform spin susceptibility $\chi_{\mu\nu}=
{\lim}_{\q \rightarrow 0} {\lim}_{\Omega_{n} \rightarrow 0}
\chi_{\mu\nu}(q)$ 
which can be decomposed into a 
Van-Vleck and Pauli part as, 
\begin{eqnarray}
&& \chi_{\mu\nu}^{\rm V}=
{\lim}_{\Omega_{n} \rightarrow 0}{\lim}_{\q \rightarrow 0} 
\chi_{\mu\nu}(q), 
\\ && 
\chi_{\mu\nu}^{\rm P}= \chi_{\mu\nu} - \chi_{\mu\nu}^{\rm V}. 
\end{eqnarray}
 We obtain the following expressions, 
\begin{eqnarray}
\label{eq:vanvleck-pauli}
&&\hspace{-10mm}
\chi_{\mu\nu}^{\rm P}=
- \lim_{\q \rightarrow 0} 
\lim_{\Omega_{n} \rightarrow 0}
\sum_{\gamma} \sum_{k}{'} 
\nonumber \\ && \hspace{-7mm}
[S^{\mu}_{\gamma\gamma}(\k,\k) S^{\nu}_{\gamma\gamma}(\k,\k)
G_{\gamma}(k+q)G_{\gamma}(k) 
\nonumber 
\\&& \hspace{-10mm}
-S^{\mu}_{\gamma\gamma}(\kp,\kp) S^{\nu}_{\gamma\gamma}(-\km,-\km)
F_{\gamma}(k+q)F^{\dag}_{\gamma}(k)],
\\&& \hspace{-10mm}
\chi_{\mu\nu}^{\rm V}=
-\sum_{\gamma \ne \delta} \sum_{k}{'} 
[S^{\mu}_{\gamma\delta}(\k,\k) S^{\nu}_{\delta\gamma}(\k,\k)
G_{\delta}(k)G_{\gamma}(k)
\nonumber 
\\&& \hspace{-3mm}
-S^{\mu}_{\gamma\delta}(\kp,\kp) S^{\nu}_{\gamma\delta}(-\km,-\km)
F_{\delta}(k)F^{\dag}_{\gamma}(k)]. 
\end{eqnarray}
 Assuming $|\Delta_{\gamma}(\k)|, v_{\rm F} |\qfflo|/2 
\ll |\alpha|$ where $v_{\rm F}$ is the Fermi velocity, 
the Van-Vleck part of spin susceptibility eq.~(A.5) is obtained as in
eq.~(16).

 When we restrict to the AFM moment along the principal axis, 
namely $\vec{h}_{\rm Q} \parallel \hat{x}, \hat{y}$ or $\hat{z}$, 
the relation $\hat{U}(-\k) = e^{{\rm i}\theta} \hat{I} \hat{U}(\k) $ holds 
with  $\theta$ an arbitrary phase factor. 
 Here we denote 
\begin{eqnarray}
\label{eq:inversion-matrix}
&&
\hat{I} =
\left(
\begin{array}{cc}
\hat{I}_{2} & 0 \\
0  & \pm \hat{I_{2}} \\
\end{array}
\right),
\\ 
&&
\hat{I}_{2} =
\left(
\begin{array}{cc}
1 & 0 \\
0  & -1 \\
\end{array}
\right). 
\end{eqnarray}
The sign of $\hat{I}_{2}$ in $ \hat{I} $ is $+$ for 
$\vec{h}_{\rm Q} \parallel \hat{x}$, $\hat{y}$ and 
$-$ for $\vec{h}_{\rm Q} \parallel \hat{z}$.
 According to eqs.~(17) and (18), we obtain 
$S^{\mu}_{\gamma\gamma}(-\k,-\k)=-S^{\mu}_{\gamma\gamma}(\k,\k)$ 
for $\mu =x,y$ and 
$S^{\rm z}_{\gamma\gamma}(-\k,-\k)=S^{\rm z}_{\gamma\gamma}(\k,\k)$. 
If $v_{\rm F} |\qfflo|/2 \ll |\alpha|$, 
the coefficient in eq.~(A.4) is approximated as  
$
S^{\mu}_{\gamma\gamma}(\kp,\kp) S^{\nu}_{\gamma\gamma}(-\km,-\km)
\sim S^{\mu}_{\gamma\gamma}(\k,\k) S^{\nu}_{\gamma\gamma}(-\k,-\k)
$
and the Pauli part of spin susceptibility is obtained as eq.~(15).

\section{magnetic Properties in the $d$-wave State}

 For the discussion for the intra-plane $d$-wave state 
we adopt the model eq.~(29) but assume the tight binding parameters 
in eq.~(27) as,  
\begin{eqnarray}
(t_1,t_4,n) = (1,0.2,0.8), 
\end{eqnarray}
with all other parameters zero. 
 This parameter set leads to the nearly half-filled band with quasi-two 
dimensional Fermi surface and leads to the dominantly 
$d_{\rm x^{2}-y^{2}}$-wave SC state for the parameter set 
(A) $U>0$, $V=-0.8U$. 
 The order parameters are described as 
$\Phi(\k) = \delta + \eta \cos k_{\rm x} - \cos k_{\rm y} $ with 
$\delta=0$ and $\eta=1$ at $h_{\rm Q}=h=0$. 
 Our analysis confirms $|\delta|, |1-\eta| \ll 1$. 
 In general, the $d_{\rm x^{2}-y^{2}}$-wave state is admixed with the 
$f_{\rm x(x^{2}-y^{2})}$- and $f_{\rm y(x^{2}-y^{2})}$-wave order parameters 
owing to the ASOC. 
 However, the $f$-wave component does not appear in the mean field solution 
of the effective model eq.~(29) because 
interactions beyond the nearest neighbor sites are neglected.

\begin{figure}[htbp]
  \begin{center}
\includegraphics[width=6.5cm]{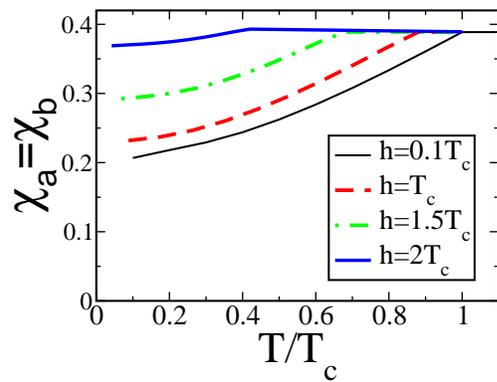}
\caption{(Color online)
The spin susceptibility along the [100]- and [010]-directions 
in the $d$-wave state without AFM order. 
We assume $U > 0$, $V=-0.8U$, $J=0$, $\alpha=0.3$ and $h_{\rm Q}=0$. 
The magnetic field is chosen as $h=0.1 T_{\rm c}$, 
$h=T_{\rm c}$, $h=1.5 T_{\rm c}$ and $h=2 T_{\rm c}$ 
from the bottom to the top. 
}
  \end{center}
\end{figure}

 We calculate the critical magnetic field \hc by solving the linearized 
mean field equation eqs.~(40) and (41) and show the result in Fig.~4. 
 The spin susceptibility is calculated on the basis of eq.~(43) by 
solving the mean field equation eqs.~(35-38). 
 In Fig.~B.1 we show the spin susceptibility below \Tc 
for various magnetic fields. 
 These results should be contrasted to those for the $p$-wave state 
(Figs.~4, 6, 10 and 11).


\begin{thebibliography}{9}
\bibitem{rf:comment}
In this sense, the recent Knight shift measurement for 
Li$_2$Pt$_{3}$B and  Li$_2$Pd$_{3}$B looks puzzling. 
Although the ASOC is much larger than the SC gap 
in both compounds,~\cite{rf:lee-pickett} 
the qualitatively different results have been obtained in 
the NMR Knight shift.~\cite{rf:zheng}
\end{thebibliography}
\end{document}